%% file: v652_sim.tex
\documentclass[usenatbib]{mnras}
\usepackage{epsfig}
\usepackage{times}
\usepackage{amsmath,amssymb}
\usepackage{xcolor}
\usepackage{tikz}
\graphicspath{{figures/}}

\input{mydefs}
\newcommand{\appropto}{\mathrel{\vcenter{
  \offinterlineskip\halign{\hfil$##$\cr
    \propto\cr\noalign{\kern2pt}\sim\cr\noalign{\kern-2pt}}}}}

\title[Spectrum synthesis for shocked stars]{Spectrum synthesis for radially pulsating stars with shocked atmospheres}

\author[C. S. Jeffery]{
  C. Simon Jeffery\thanks{E-mail:simon.jeffery@armagh.ac.uk} 
 \\
Armagh Observatory and Planetarium, College Hill, Armagh BT61 9DG, N. Ireland, UK
 }
 
\date{Accepted \ldots; Received \ldots; in original form \ldots}
\pagerange{\pageref{firstpage}--\pageref{lastpage}} 
\pubyear{2021} 

\begin{document}
\label{firstpage}
\maketitle

\begin{abstract} {\sc spec\_puls} describes a suite of computer programs to simulate the emergent spectrum from a radially-pulsating star. 
It combines a Christy-type non-linear pulsation code with  classical stellar atmosphere codes. 
The principal aim is to interpret the dynamical spectrum of the radially pulsating extreme helium star V652\,Her, which shows a strong shock at minimum radius. 
The components are general enough to treat other classes of radial pulsation. 
The theoretical spectrum from a shocked pulsation model shows line doubling, with the blue component emerging at standstill velocity and accelerating blueward.  
The doubling phase depends on line depth and parent ion. 
The behaviour of line cores post-shock points to a drop in the ionization temperature, although the gas temperature in the model remains high.
Shock compression leads to phase-dependent strengthening of Stark-broadened line wings, with the far wings responding first. 
With velocity, temperature and pressure-sensitive diagnostics, detailed tomography of the pulsation-driven shock in V652\,Her seems possible. 
Even when no shock is present, the dynamical spectrum is significantly different from a model in hydrostatic equilibrium. 
Using the quasi-static approximation (e.g. at maximum radius) may lead to a considerable underestimate of the star's mean effective temperature and surface gravity.
\end{abstract}

\begin{keywords}
hydrodynamics
-- line profiles
-- radiative transfer
-- shock waves
-- stars: oscillations
-- stars: individual: V652\,Her
\end{keywords}

\section{Introduction}

Large-scale motions play a significant r\^ole in both the structure of and emergent spectrum from the atmosphere of a pulsating star, particularly when the pulsation is of sufficient amplitude that a shock is formed at minimum radius. 

For radial pulsations, it is well-known that the isolated absorption line formed during expansion or contraction phases must display an asymmetric profile. 
Simply consider that, from the point of view of a distant observer, a larger contribution from Doppler-shifted absorption due to vertical motion at the disk centre will combine with a weaker contribution from unshifted absorption due to the absence of tangential motion near the stellar limb. 
This profile will vary around the pulsation cycle from a blue-shifted asymmetric profile in the expanding phase  through a symmetric profile at maximum radius to a red-shifted asymmetric profile during contraction \citep[e.g.][]{parsons72, albrow94, montanes01}.

Less obvious is the behaviour of the line profile at minimum radius.
At this phase, in-falling material from the outermost layers of the star meets material from lower layers being driven outwards by the next pulsation cycle. 
In some cases, radially pulsating stars appear to show double line profiles at minimum radius, a phenomenon first attributed to a shock wave by \citet{schwarzschild52} and \citet{whitney56}. 
 Examples include RR\,Lyrae and W\,Virginis stars \citep{sanford49, wallerstein59}, long period variables \citep{joy47}, $\beta$\,Cepheid variables \citep{odgers56}, and the R Coronae Borealis star RY\,Sgr \citep{lawson86}.

The velocity curve of the radially pulsating extreme helium star V652\,Herculis (V652 Her) shows a substantial jump at minimum radius \citep{hill81,lynasgray84,jeffery86,jeffery01b}.
Further improving the temporal resolution, \citet{jeffery15b} demonstrated a discontinuity in line position, the phase of which is in inverse proportion to the depth at which the line is formed. 
The conclusion from both observations and non-linear pulsation models \citep{fadeyev96,montanes02,jeffery22a} is that a strong shock runs through the photosphere as the surface of the star is accelerated outwards. 
In contrast, radial pulsations in the very similar extreme helium star BX\,Circini (BX\,Cir) show a smoother radial velocity curve \citep{kilkenny99,woolf02}, and no evidence of line doubling \citep{martin19.phd}. 

The na\"ive logic is that the collision between inward and outward layers inevitably leads to compression and, at least, adiabatic heating. 
At the resolution of hydrodynamic models, a "jump" in the line velocities is produced when the pulsation is of sufficient amplitude \citep{jeffery22a}.

Armed with superlative spectroscopic data as well as detailed non-linear pulsation models covering the pulsation cycle of V652\,Her (and BX\,Cir), our goals were to develop an algorithm for modelling the spectrum of a radially pulsating helium star, with particular attention to phases where the atmosphere is shocked, and to establish how the \citet{schwarzschild52} scenario affects the spectrum in these cases.
The question has been addressed for shocks in other pulsating stars by, for example, \citet{sasselov92} for classical Cepheids, and \citet{fokin04} for $\beta$ Cepheids. 
\citet{alvarez00} developed a tomographic method to measure velocity gradients in the atmospheres of long-period variables, essentially by combining line-profile information from lines formed at different depths, and hence demonstrated the presence of shocks. 
A similar approach was adopted for V652\,Her by \citet{jeffery15b}. 
Since the physical conditions in V652\,Her differ substantially from any of these examples, not least because the low hydrogen abundances increases the transparency of the atmosphere, such models have a predictive r\^ole in exploring what might be observed, as well as an interpretative r\^ole in understanding what has been observed. 

\citet{jeffery13.fuji2} included a brief description of our method. Having presented the non-linear pulsation models \citep{jeffery22a}, this paper expands the spectral synthesis calculations.  
In \S~\ref{s:theory} we treat the radiative transfer for an equilibrium model atmosphere perturbed to match the density, temperature and velocity structure provided by a non-linear radial pulsation model.  
We adopt 
(i) a LTE plane-parallel code to compute the structure of non-grey model atmospheres  (\S~\ref{s:sterne}), 
(ii) an adaptation of our code for computing the formal solution of the radiative transfer equation  (\S~\ref{s:formal}), 
(iii) the nonlinear hydrodynamic code  for modelling the pulsations  (\S~\ref{s:hydro}), and 
(iv) an algorithm to combine this information and yield the emergent spectrum as a function of pulsation phase (\S~\ref{s:puls}).
A test calculation is described in \S\,\ref{s:tests}.
A description of the structure of the dynamical atmosphere is given in \S~\ref{s:shock}. 
Examples of line profiles predicted by various pulsation models are given in \S~\ref{s:examples}. 
Implications for V652\,Her, BX\,Cir and other stars are discussed in \S~\ref{s:discussion}.
\S~\ref{s:conclusion} concludes.  
Access to copies of \citet{jeffery15b} and \citet{jeffery22a} would assist the reader. 

\section{Theory}
\label{s:theory}

\subsection{Plane-parallel model atmospheres: {\sc sterne}}
\label{s:sterne}

Our starting approximation is that the star is spherically symmetric and the surface layers responsible for the emergent spectrum (atmosphere) are thin compared with the radius of the star. Hence we adopt the plane-parallel approximation. We assume that the macroscopic and atomic properties of the atmosphere, 
when stationary, are in local thermodynamic, hydrostatic and radiative equilibrium. 
We also assume the atmosphere to be chemically homogeneous, though this can be relaxed.
We then compute model atmospheres for a given effective temperature (\Teff), surface gravity ($g$) and chemical composition($X_i$, where $i$ represents atomic number) using the code {\sc sterne} \citep{behara06}. 
Continuous opacities $\kappa_c$ are calculated from Opacity Project photo-ionization cross sections \citep{OP94,OP95}; 
line opacities $\kappa_l$ are treated using the opacity sampling approach, currently including the contribution of $\approx 900,000$  lines \citep{behara06}. 
$\sigma$ is the  electron-scattering coefficient. 
Line broadening incorporates a microturbulent velocity $v_{\rm t}$.   
The output is a numerical description of the atmosphere in terms of temperature $T_{\star}$, pressure $P_{\star}$ and density $\rho_{\star}$ as functions of optical depth $\tau$:
\begin{equation}
T_{\star}(\tau), P_{\star}(\tau), \rho_{\star}(\tau), 
\end{equation}
where $\tau$ is the continuum optical depth at a reference wavelength (usually 4000\AA\ in our calculations).  
We integrate the identities 
\begin{equation}
{\rm d}z = \frac{{\rm d} \tau}{(\kappa+\sigma)}, {\rm d}m' = \rho {\rm d}z
\end{equation}
to obtain the geometric depth $z(\tau)$ and column mass $m(\tau)$  and invert to obtain  
\begin{equation}
z(m'), \tau(m'). 
\end{equation}

\begin{figure}
\begin{center}
\begin{tikzpicture}
\draw (0,4) arc (90:30:4);
\draw (0,4) arc (90:150:4);
\draw (0,4.5) arc (90:30:4.5);
\draw (0,4.5) arc (90:150:4.5);
\draw [->] (-3,2.3) -- (-3,5.5);
\draw [->] (-1.5,3.2) -- (-1.5,5.5);
\draw [->] (0,3.5) -- (0,5.5); 
\draw [->] (1.5,3.2) -- (1.5,5.5) node[anchor=south] {$I_{\nu\mu}$};
\draw [->] (3,2.3) -- (3,5.5);
\draw (1.5,4.243) -- (1.8,5.091) node[anchor=south] {$\mu$};
\draw (1.5,4.9) arc (72.17:69.29:5.12);
\draw [very thick,->] (2.25,3.897) -- (2.10,3.626);
\draw [very thick,->] (2,3.464) -- (1.8,3.118) node[anchor=north] {$\dot{z}(\tau)$};
\end{tikzpicture}
\end{center}
\caption{Schematic of the emergent flux from a differentially moving stellar atmosphere. 
The vertical vectors representing the line of sight to the observer indicate how the specific intensities $I_{\nu\mu}$ are integrated over optical depth $\tau$. 
Radial vectors refer to the radial velocity $\dot{z}(\tau)$. 
The emergent flux $F_{\nu}$ is the integral of $I$ over $\mu$. 
}
\label{f:geom}
\end{figure}
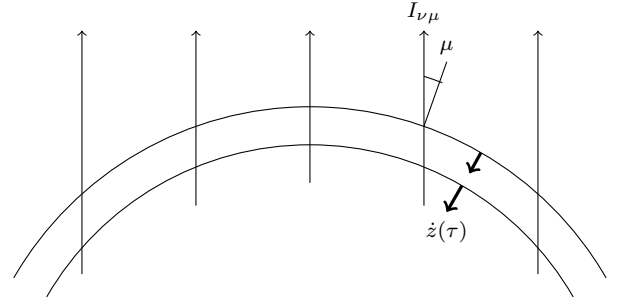

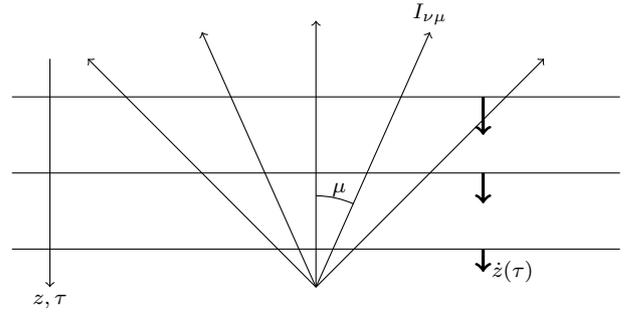
\begin{figure}
\begin{center}
\begin{tikzpicture}
\draw (-4,3) -- (4,3);
\draw (-4,2) -- (4,2);
\draw (-4,1) -- (4,1);
\draw [->] (-3.5,3.5) -- (-3.5,0.5) node[anchor=north] {$z,\tau$};
\draw [->] (0,0.5) -- (-3,3.5); 
\draw [->] (0,0.5) -- (-1.5,3.85); 
\draw [->] (0,0.5) -- (0,4); 
\draw [->] (0,0.5) -- (1.5,3.85) node[anchor=south] {$I_{\nu\mu}$}; 
\draw [->] (0,0.5) -- (3,3.5); 
\draw (0,1.7) arc (90:65.55:1.2) node[anchor=south east] {$\mu$}; 
\draw [very thick, ->] (2.2,1) -- (2.2, 0.7) node[anchor=west] {$\dot{z}(\tau)$}; 
\draw [very thick, ->] (2.2,2) -- (2.2, 1.6); 
\draw [very thick, ->] (2.2,3) -- (2.2, 2.5); 
\end{tikzpicture}
\end{center}
\caption{The schematic of Fig.\,\ref{f:geom} represented in plane-parallel geometry.
}
\label{f:plane}
\end{figure}

\subsection{Formal solution: {\sc spectrum}}
\label{s:formal}

For a plane-parallel stellar atmosphere, the emergent flux at frequency $\nu$ is defined by the standard integral over angle $\mu = \cos \theta$ of the specific intensity $I_{\nu\mu}$:
\begin{equation}
 F_{\nu} = 2\pi  \int_{-1}^{1}  I_{\nu\mu} \mu d\mu
\label{eq_flux} 
\end{equation}
The specific intensity is obtained from the integral of the source function
\begin{equation}
I_{\nu\mu} = \int_{0}^{\infty} S_{\tau\nu} \exp (-\tau/\mu)
d\tau / \mu
\label{eq_intens}
\end{equation}
over optical depth $\tau$.

In the case of an atmosphere moving with uniform velocity $\dot{z}$
it is sufficient to treat $S_{\tau\nu}$ as independent of $\mu$. 
The integral over angle (Eqn.\,\ref{eq_flux}) needs only to consider the projected specific intensity by introducing the appropriate $\tau$-independent frequency shift
\begin{equation}
\nu_{\mu} = \nu_{0} ( 1 + \frac{\dot{z}\mu}{c} ).
\label{eq_freq1}
\end{equation}
Eqns.\,\ref{eq_flux} -- \ref{eq_freq1} are then sufficient to reproduce the
line profile through the bulk of the pulsation cycle
\citep[cf.][]{montanes01}.  

However, if radial velocity is a function of optical depth $v(\tau)$ (Figs. \ref{f:geom} and \ref{f:plane}), then 
\begin{equation}
\nu_{\tau\mu} = \nu_{0} ( 1 +  \frac{\dot{z}(\tau)\mu}{ c } )
\label{eq_freq}
\end{equation}
and Eqn.\,\ref{eq_intens} must treat the frequency dependence of $S$ explicitly. 
The source function is conventionally defined by
\begin{equation}
S_{\tau\nu} =
(\kappa_{\tau\nu}B_{\tau\nu}+\sigma_{\tau}J_{\tau\nu})/(\kappa_{\tau\nu}+\sigma_{\tau})
 \label{eq_source}
\end{equation}
where $B_{\tau\nu}$ is the Planck function, the mean intensity $J_{\tau\nu}$ is found by solution of the transfer equation, the scattering coefficient  $\sigma_\tau$ is normally independent of frequency,  and  the total absorption coefficient $\kappa$ is the sum of the continuous absorption  coefficient $\kappa_{c}$, which varies slowly with $\nu$, and the line absorption coefficient  $\kappa_{l}$. 
 Since strict local thermodynamic equilibrium (LTE) requires $S_{\nu} = B_{\nu}$, the inclusion of scattering implies a partial non-LTE approach.
\begin{equation}
\kappa_{\nu\mu\tau} = \kappa_{l\nu\mu\tau} + \kappa_{c\nu\tau} 
\label{eq_ktotal}
\end{equation}
We make the approximation that  $\kappa$ is independent of $\mu$ ({\it i.e.} complete redistribution).

In the case that $v(\tau)$ varies with depth, the specific intensities must be computed by integrating Eqn.\,\ref{eq_intens} with  
\begin{equation}
S_{\tau\nu\mu} =
(\kappa_{\tau\nu\mu}B_{\tau\nu}+\sigma_{\tau}J_{\tau\nu})/(\kappa_{\tau\nu\mu}+\sigma_{\tau})
\label{eq_smove}
\end{equation}
now calculated from the correctly projected local line opacity  
\begin{equation}
\kappa_{l\tau\mu}(\nu) = \kappa_{l\tau\mu}(\nu_0.(1 + \frac{\dot{z}(\tau)\mu}{c})).
\label{eq_kline}
\end{equation}

These equations have been implemented within the LTE formal solution code {\sc spectrum} \citep{jeffery01b}, which computes theoretical line profiles and spectra for a given model atmosphere structure. 
To the description of temperature and pressure as a function of optical depth provided by conventional model atmospheres, we add a description of the vertical  velocity structure $\dot{z}(\tau)$. 
Since $\tau$ increases with depth into the atmosphere,  $\dot{z}(\tau) > 0$ corresponds to a contracting layer and hence red-shifted absorption as seen by a distant observer. $\dot{z}(\tau) < 0$ corresponds to an expansion and blue-shifted absorption. 
The plane-parallel approximation is justified because sphericity effects in the stars under investigation here are generally small, weakening the line but not influencing its shape.

Here we adopt an input model atmosphere computed assuming hydrostatic, radiative and local thermal equilibrium, but including scattering. 
For the slowly varying phases of a star's pulsation cycle, these approximations are probably satisfactory except in the case of large amplitude pulsations in very luminous stars. 

\subsection{Hydrodynamic models for stellar pulsation: {\sc nlpuls}}
\label{s:hydro}

The non-linear code {\sc nlpuls} is a Christy-type code \citep{christy67} written originally by \citet{bridger83} and substantially enhanced by \citet{montanes02}  to include OPAL or OP opacities and much finer mass zoning. 
\citet{jeffery22a} explored a large volume of parameter space pertinent to the pulsations observed in V652\,Her. 
A pulsation model is defined by the principal parameters stellar mass $M$, effective temperature $\Teff$,  luminosity $L$ and chemical composition $X_i$. The stellar radius $R$ is implicit.   
Here, we use the labels $llttmmXX$ adopted by \citet{jeffery22a}, where 
$ll = {\rm frac}({ \log L/\Lsolar})$, $tt = {\rm frac}({\log \Teff})$, $mm = {\rm frac}({M/\Msolar})$ and $XX$ is a label representing chemical composition and opacity, and where ${\rm frac}(x)$ represents the fractional or decimal part of a non-negative real-number $x$.  

A prior for computing the {\it dynamical} model is a static envelope and atmosphere in hydrostatic equilibrium, which is characterised as a function of mass by
\begin{equation}
r_0(m), T_0(m), P_0(m), (\dot{r}_0\equiv 0),
\end{equation} 
where $m=0$ at the centre of the star and $m=M$ at the surface and  the variables represent radius $r$, pressure $P$, temperature $T$,  and radial velocity $\dot{r}$. 
The inner boundary of the model is at roughly 1/10 the stellar radius ($R/10$). 
The outer boundary extends into the outer atmosphere.

Each {\it dynamical} model describes the behaviour of these variables as a function of time $t$ or pulsation phase $\phi$: 
\begin{equation}
r(m,\phi), T(m,\phi), P(m,\phi), \dot{r}(m,\phi).
\end{equation}

\subsection{Line profiles and spectra for pulsating stars: {\sc spec\_puls}} 
\label{s:puls}

Dropping arguments, we parameterise the pulsation in terms of scaled variables 
\begin{equation}
r' = r/r_0, T' = T/T_0, P' = P/P_0.
\end{equation}
Rewriting the mass variable
\begin{equation}
m' = (M - m) / 4\pi r^2
\end{equation}
allows the scaled dynamical model to be mapped onto an arbitrary hydrostatic plane-parallel model atmosphere. 

 A series of pulsation phases $\phi_i$ is selected, the equilibrium atmosphere variables $P_{\star}(m')$ and $T_{\star}(m')$ are multiplied by $P'(m',\phi)$ and $T'(m',\phi)$, and the local velocity distribution $\dot{r}(m',\phi)$ is introduced exactly. 
Eqns.\,11, 10, 5 and 4 are then evaluated to provide the flux $F_{\nu}(\phi)$ at each $\phi_i$. 

Our notation distinguishes geometric depth $z$ from the top of the atmosphere and radial distance $r$ from the centre of a star of radius $R$. In general $z \neq R - r$, since $R$ is usually defined at some finite optical (e.g. $\tau = 2/3$), but we can equate velocities
\begin{equation}
 \dot{z}(\tau) \equiv \dot{z}(m') = -\dot{r}(m').  
\end{equation}
Neither quantity should be confused with apparent red- or blue-shift expressed in velocity units $v$  discussed in subsequent sections.

\begin{figure}
\begin{center}
\begin{tikzpicture}
\draw[thick,->] (0,0) -- (7.0,0) node[anchor=west] {$\log \tau$};
\draw[thick,->] (0,1.5) -- (0,-1.5) node[anchor=north] {$\dot{z}(\tau)$};
\draw (0,-1) -- (2,-1) -- (5,1) -- (7,1);  
\draw (0,-1) -- (2,-1) node[anchor=north] {in-falling};
\draw (7,1) -- (5,1) node[anchor=south] {out-flowing};
\draw (0,0) -- (.1,0) node[anchor=south west] {top};
\draw (7,0) -- (6.9,0) node[anchor=south east] {bottom};
\draw (3.5 cm,-2pt) -- (3.5 cm,2 pt) node[anchor=north west] {$\tau_0$};
\draw (2pt,1 cm) -- (0,1 cm) node[anchor=east] {$\dot{z}(\infty)$};
\draw (2pt,-1 cm) -- (0,-1 cm) node[anchor=east]{$\dot{z}(0)$};
\draw (2 cm,-2pt) -- (2 cm,2pt) node[anchor=north]{$\tau_{\rm in}$};
\draw (5 cm,-2pt) -- (5 cm,2pt) node[anchor=north]{$\tau_{\rm out}$};
\end{tikzpicture}
\end{center}
\caption{Schematic of a test  velocity field $\dot{z}(\tau)$ applied to a  model atmosphere, showing quantities described in the text. 
In a pulsating star passing through minimum radius, the stationary point $\tau_0$ moves outwards, {\it i.e.} towards smaller optical depth. The top and bottom of the atmosphere and the direction of flow are indicated. 
}
\label{f:pert}
\end{figure}
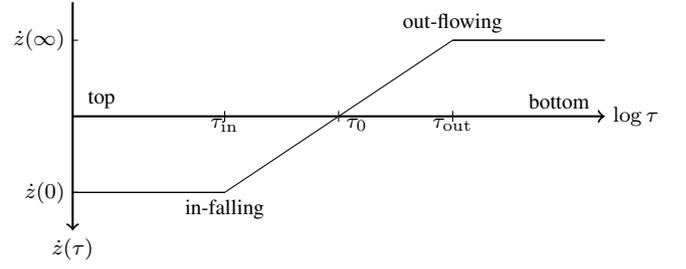

\begin{figure}
\includegraphics[width=88mm,angle=0]{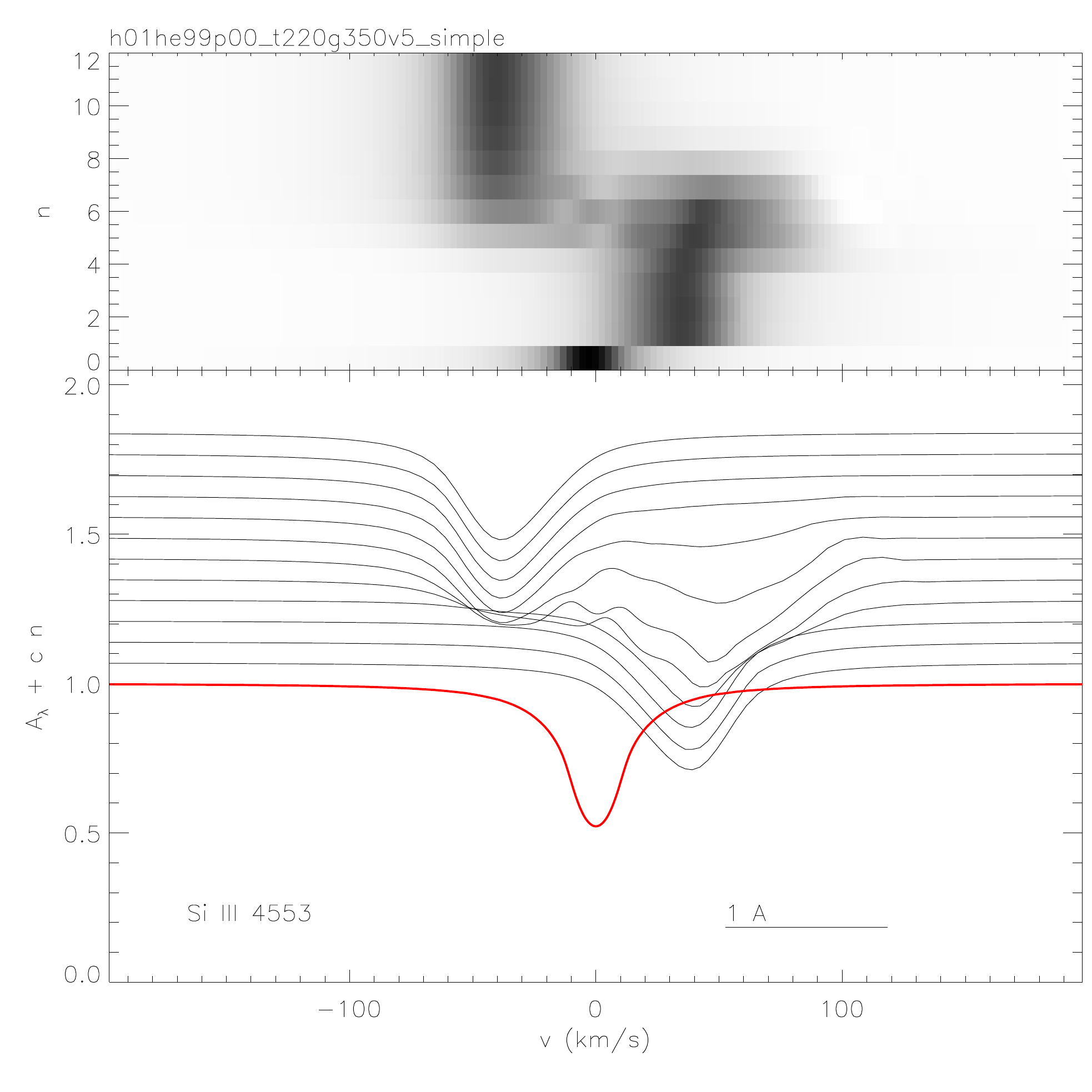}
\caption{Lower panel shows the normalised line profile for Si{\sc iii}4553 computed from a test model atmosphere in which the local velocity $\dot{z}(\tau)$ is specified, with  $\tau_0$ running from $\infty$ (second bottom: uniform contraction) through to 0 (top: uniform expansion) and $\dot{z}(0)=-\dot{z}(\infty)=-100$\,km\,s$^{-1}$ (as in Fig.\,\ref{f:pert}). The extent of the transition layer is given by $\alpha=1.2$. The line profile for a stationary atmosphere is shown in red. The upper panel shows the same data as a grey-scale image; $n$ represents the number of the spectrum in the sequence, with 0 being the stationary profile.  
}
\label{f:simple}
\end{figure}

\begin{figure*}
\includegraphics[width=88mm,angle=0,clip=true]{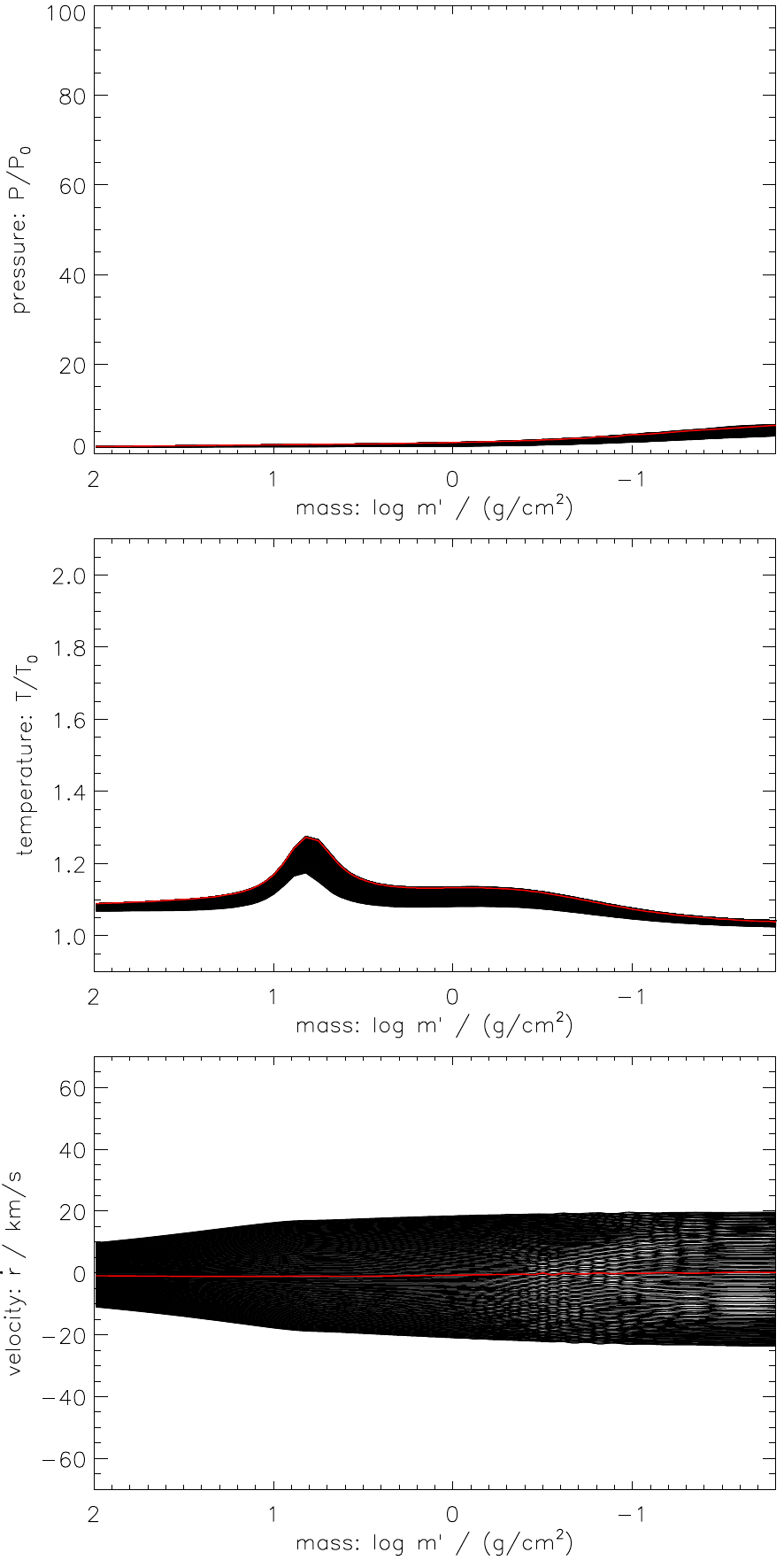}
\includegraphics[width=88mm,angle=0,clip=true]{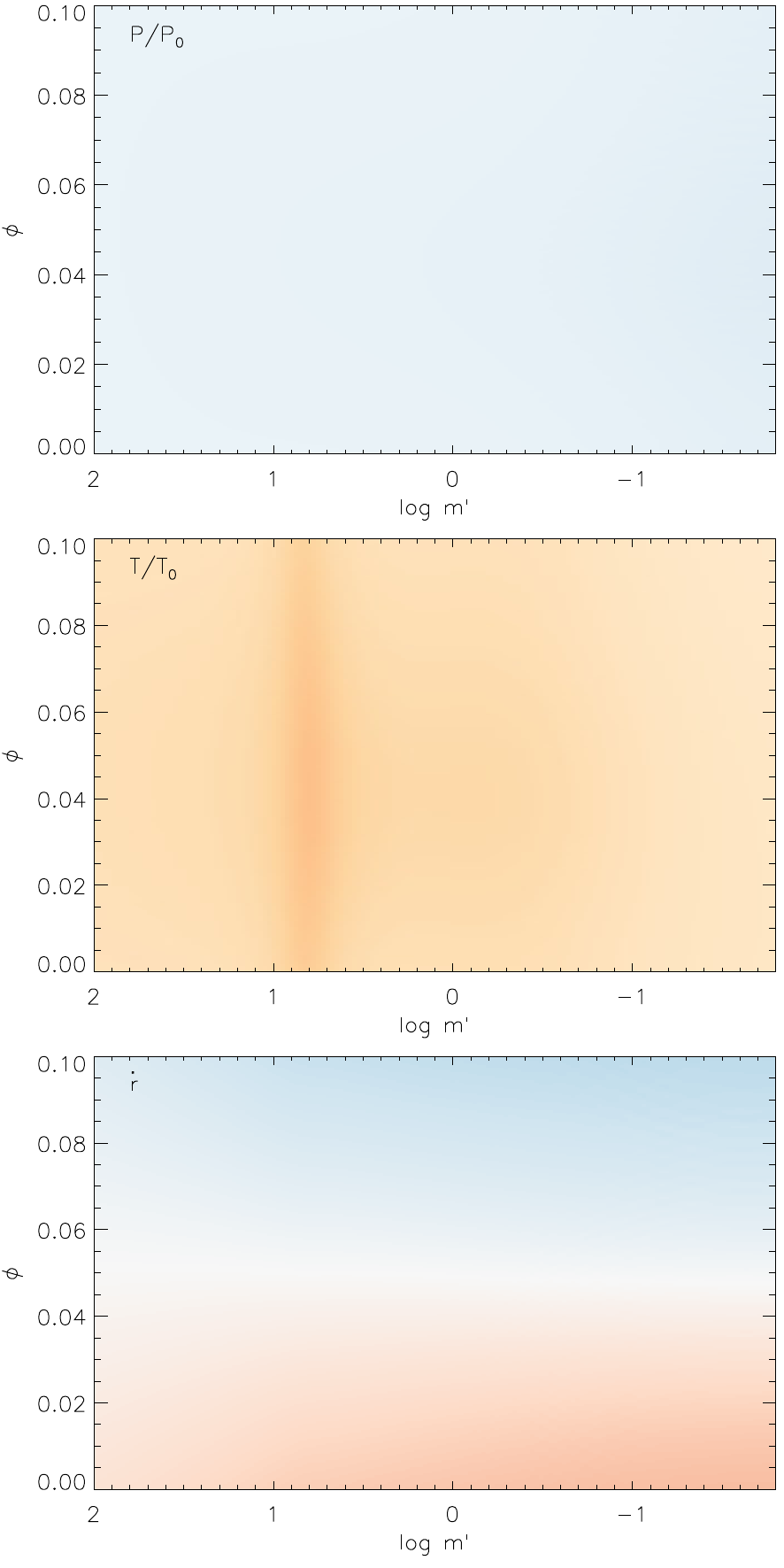}
\caption{The run of pressure (top), temperature (middle) and velocity (bottom) as functions of mass depth $m$ and pulsation phase $\phi$ for the non-shocked pulsation model 813466N1. Pressure and temperature have been scaled to their values at the same mass points in hydrostatic equilibrium. On the left, variables are shown for values of $\phi = 0 - 0.1$ lying either side of minimum radius (red). On the right, variables are shown as a colour map with $\phi$ covering the same range as on the left; the more intense, the larger the absolute value. Negative and positive velocities in the stellar centre of mass frame are represented by red and blue respectively. All scales for this figure and Fig.\,\ref{f:rmin_963666} are identical.  }
\label{f:rmin_813466}
\end{figure*}

\begin{figure*}
\begin{center}
\includegraphics[width=88mm,angle=0,clip=true]{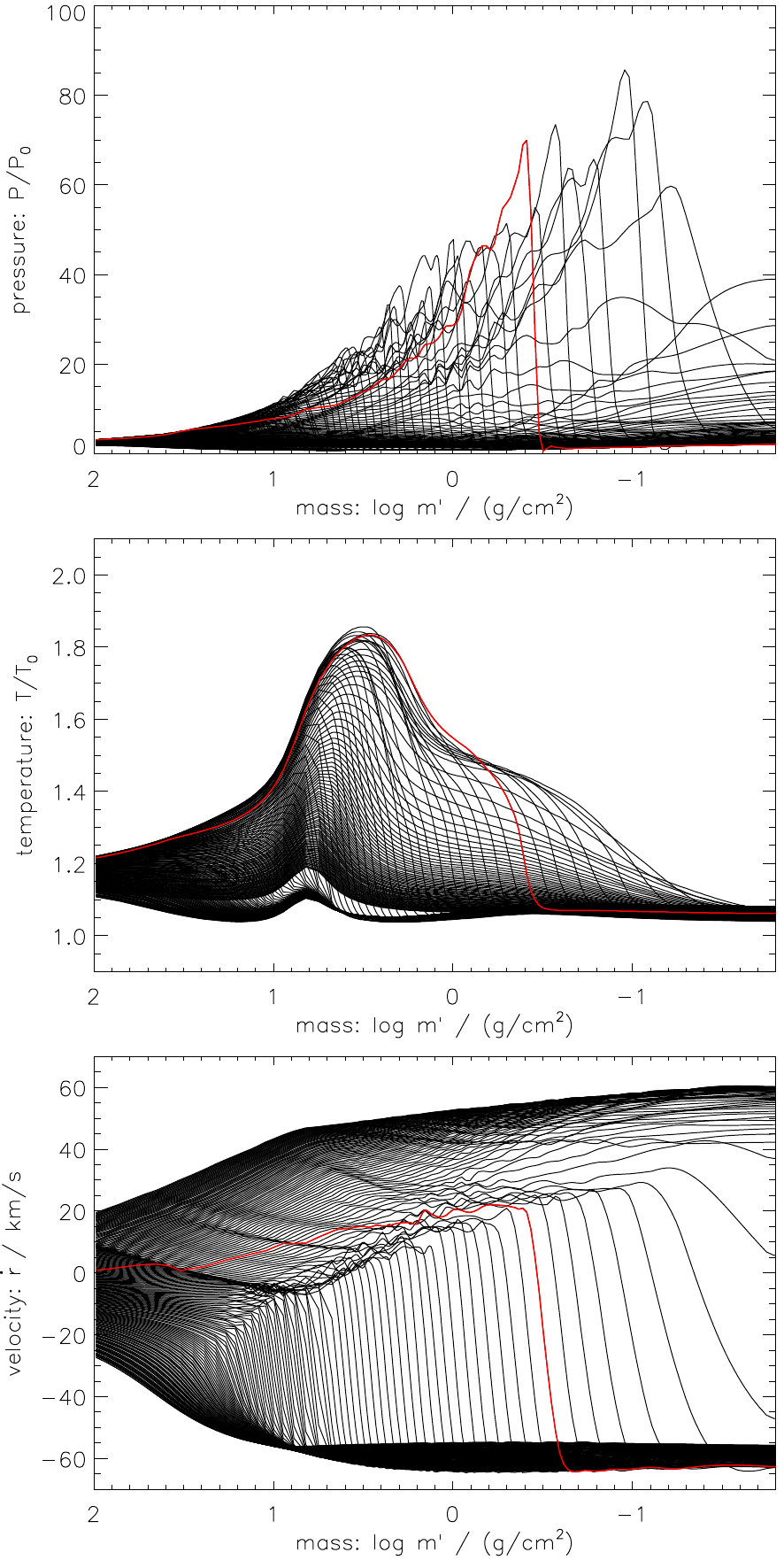}
\includegraphics[width=88mm,angle=0,clip=true]{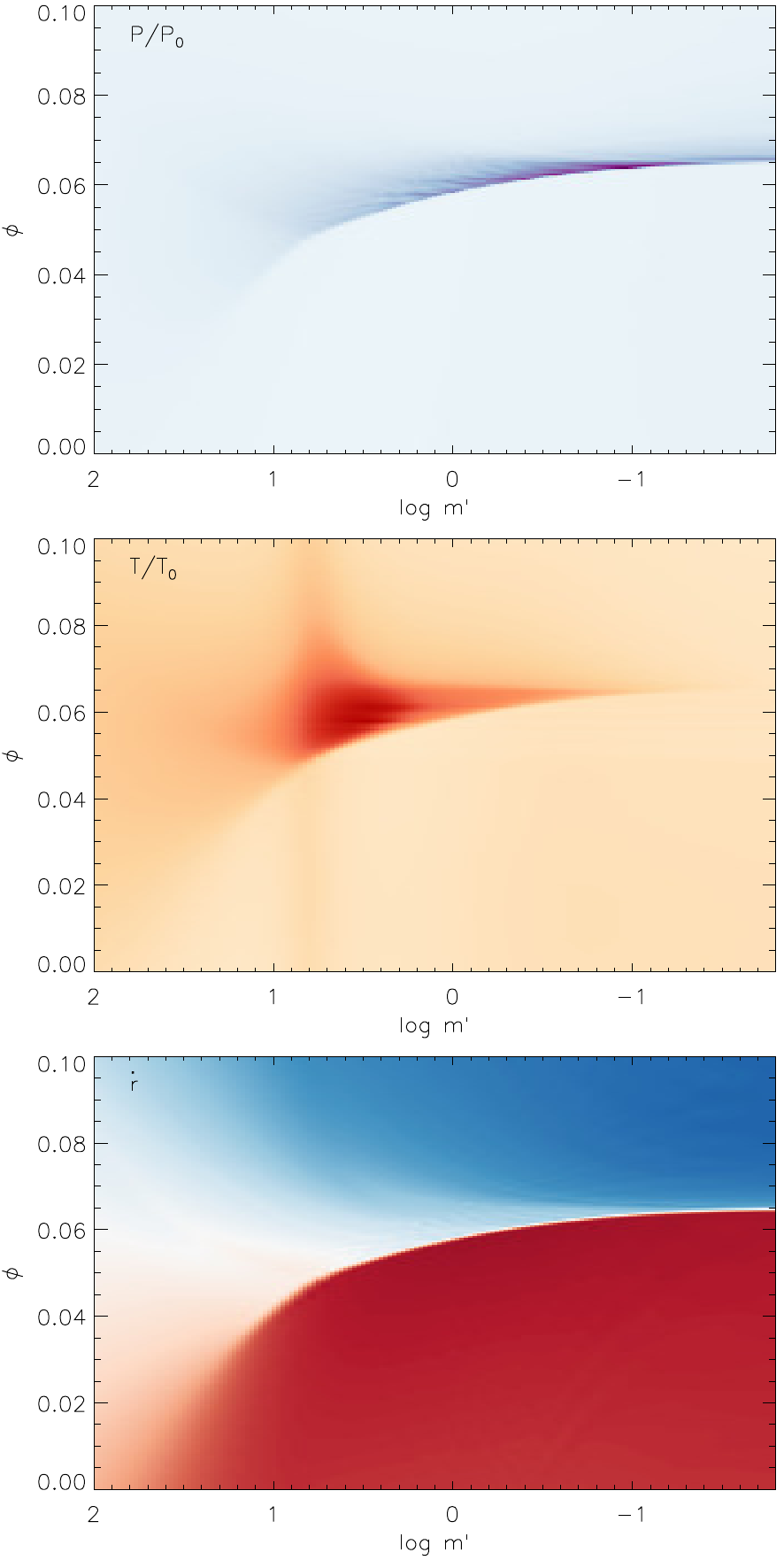}
\end{center}
\caption{As Fig.~\ref{f:rmin_813466} for the shocked pulsation model 963666N1}
\label{f:rmin_963666}
\end{figure*}

\section{Test calculations}
\label{s:tests}

In order to test the formal solution algorithm, we provided a mechanism within {\sc spectrum} to introduce i) an arbitrary radial velocity shift $v(\tau)$ and ii) an arbitrary temperature perturbation $\delta T(\tau)$ to a converged {\sc sterne} model. 

Examples of trial temperature perturbations and velocity shifts are shown in Fig.~\ref{f:pert}. 
Interesting velocity fields  involve both in-falling material and out-flowing material, normally converging to an intermediate stationary point at a specified optical depth  $\tau_0$ (Fig.\,\ref{f:pert}).  
The extent of the transition layer between in-falling material with $\tau<\tau_{\rm in}$ and  out-flowing material with  $\tau>\tau_{\rm out}$ has been parameterised using 
\begin{equation}
\alpha = \tau_{0}/\tau_{\rm in} = \tau_{\rm out}/\tau_{0}.
\end{equation}
$\alpha=1$ corresponds to a discontinuity and increasing $\alpha$ corresponds to increasing the extent of the transition layer.

Numerical problems occur when very narrow temperature perturbations are introduced into the model. 
The quadrature of Eqn.\,\ref{eq_intens} is sensitive to discontinuities in either $\kappa_{l}$ (Eqn.\,\ref{eq_kline}) or $B_{\tau\nu}$. 
The reason lies primarily in the calculation of $I_{\nu\mu}$ at  $\mu \simle 0.1$, where the integrand in Eqn.\,\ref{eq_intens} is very sharply peaked in $\tau$. 

A similar problem arises when the velocity gradient in the atmosphere is large and the calculation involves narrow (unsaturated) lines. 
This is due to the line being shifted so far in neighbouring layers that the line source function is not adequately sampled by the quadrature over $\mu$. It may arise, for example, when the velocity shift between neighbouring layers is greater than or comparable with the intrinsic line width in those layers. 

Both problems can be mitigated (but not always resolved) by a) increasing the number of optical depth  points used in the quadrature of  Eqn.\,\ref{eq_intens} and b) increasing the number of quadrature points used to evaluate Eqn.\,\ref{eq_flux}. 
Typically {\sc spectrum} uses 25 depth points and 5 lines of sight. 
For non-stationary atmospheres we increased the number of depth points by a factor 8 in the range $-3 < \log \tau < 0.5$, and used 11 lines of sight in the angle quadrature ($\mu = 0.005, 0.1, 0.2, \ldots, 1.0$). 

 Fig.~\ref{f:simple} shows a sample test calculation for an isolated absorption line in a model where the local velocity is varied from uniform contraction
 to uniform expansion in a series of steps described by varying $\tau_0$ in Fig.~\ref{f:pert} from $\infty$ to zero. The asymmetric red-shifted profile corresponding to uniform contraction transmutes to an asymmetric blue-shifted profile for uniform expansion. In between are phases in which both inward and outward moving regions of the atmosphere absorb radiation, at times producing an apparently double line. The behaviour is at least qualitatively as expected.

\begin{figure*}
\begin{center}
\includegraphics[width=88mm,angle=0,clip=true]{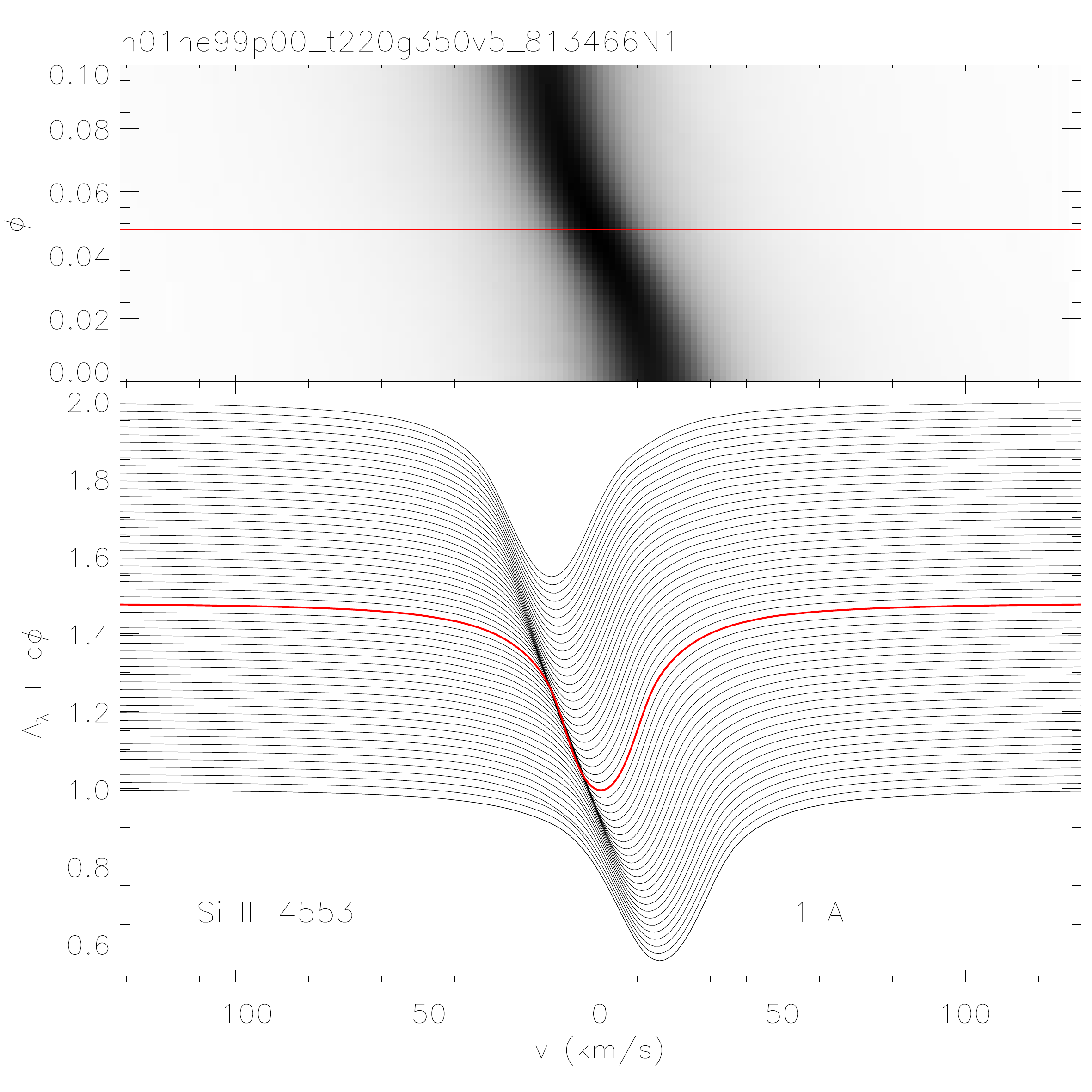}
\includegraphics[width=88mm,angle=0,clip=true]{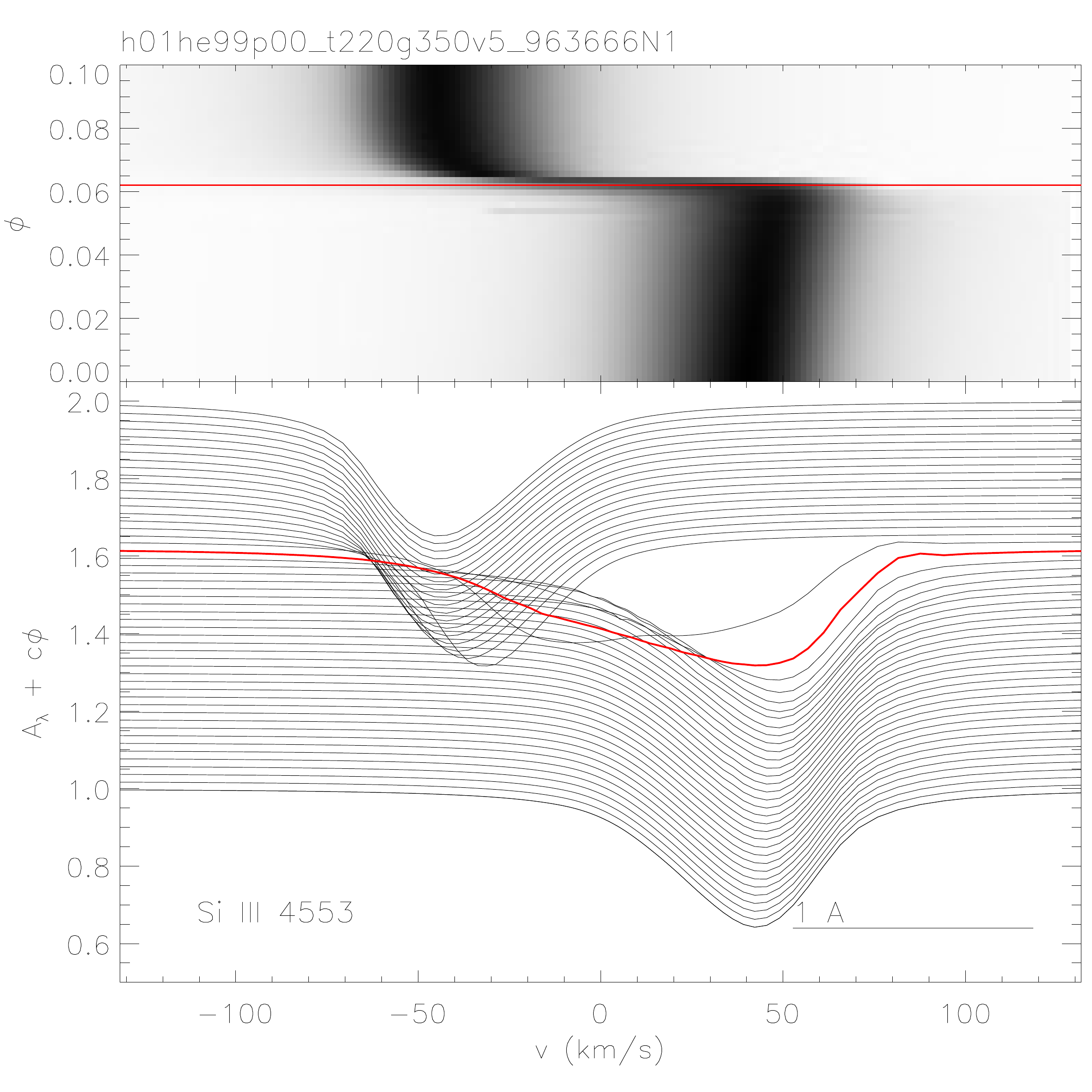}
\end{center}
\caption{Theoretical lines profiles for Si{\sc iii}\,4553\,\AA\ 
through minimum radius for two non-linear pulsation models both having  $M=0.66\,{\rm M_{\odot}}$, $X=0.00125, Z=0.0159, Y\equiv1-X-Z$, and otherwise 
$\log L/{\rm L_{\odot}} =3.81$, $\log T_{\rm eff}/{\rm K}=4.34$ (left: {\sc nlpuls} 813466N1) 
and 
$\log L/{\rm L_{\odot}} =3.96$, $\log T_{\rm eff}/{\rm K}=4.36$  (right: {\sc nlpuls} 963666N1) scaled and superposed on a LTE model atmosphere having $T_{\rm eff}/{\rm K}=22\,000$ and  $\log g/{\rm cm\,s^{-2}} = 3.5 $ and ({\sc sterne} h01he99p00\_t220g350v5). 
The lower part of each panel shows the line profiles normalised to the continuum 
($A(\phi,\lambda) = F(\phi,\lambda)/F_{\rm c}(\phi,\lambda)$)  and offset  upwards by an amount proportional to the pulsation phase which increases from 0 to 0.1 in steps of 0.002 cycles. The line profile corresponding to minimum radius is highlighted in red. Phase 0 (bottom) corresponds to maximum light in the pulsation models. Profiles are plotted in velocity units ($v$) relative to line centre in the observer's rest frame; the corresponding wavelength interval of 1\AA\ is indicated by a horizontal bar.  
The upper part of each panel shows the same information as a grey-scale image, with the continuum being represented by white and the the deepest part of the strongest line being black. 
The layout of these figures has been chosen to be comparable with figures showing the behaviour of spectral lines in the pulsating helium star V652\,Her presented by \citet{jeffery15b}.
}
\label{f:shock_comp}
\end{figure*}

\begin{figure*}
\begin{center}
\includegraphics[width=88mm,angle=0,clip=true]{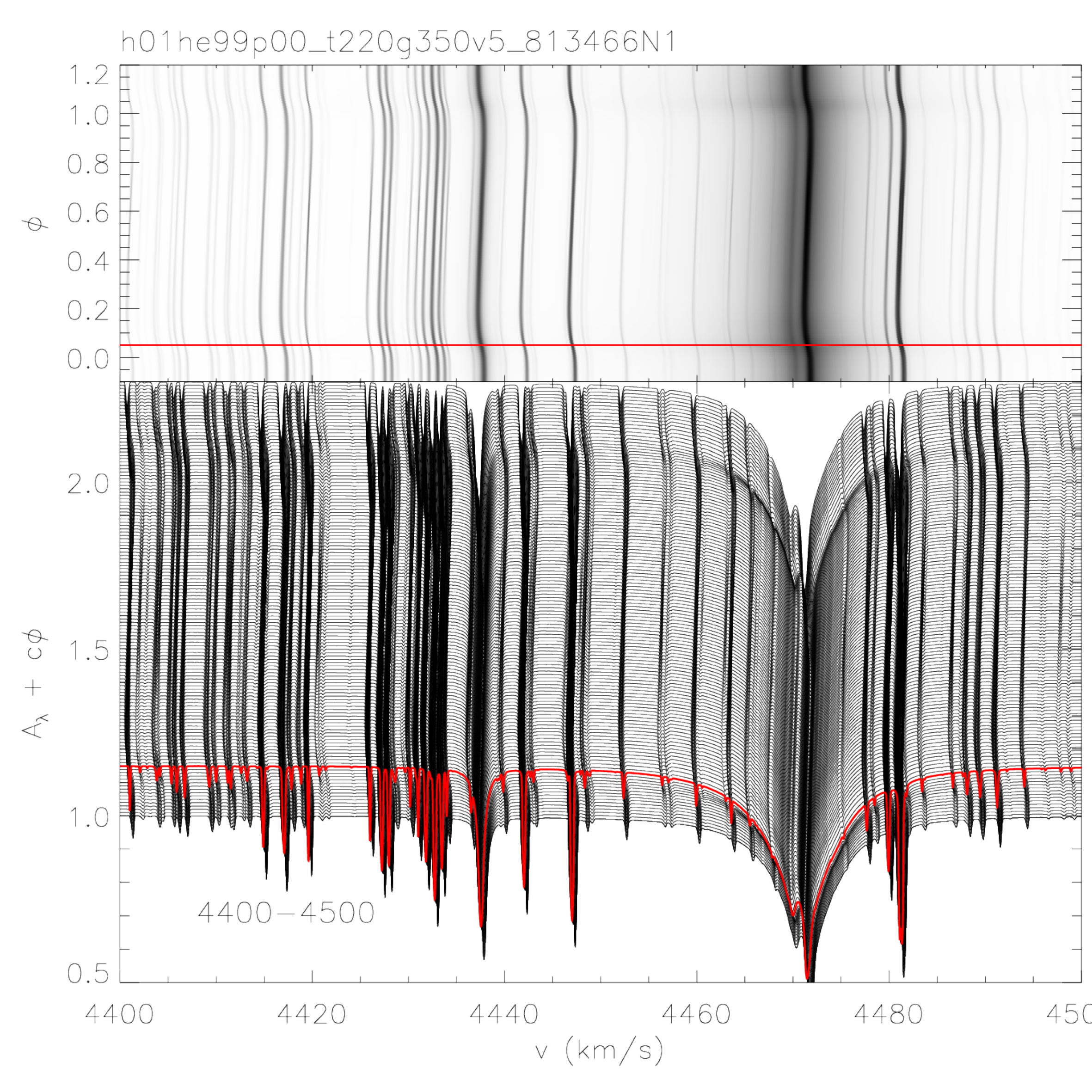}
\includegraphics[width=88mm,angle=0,clip=true]{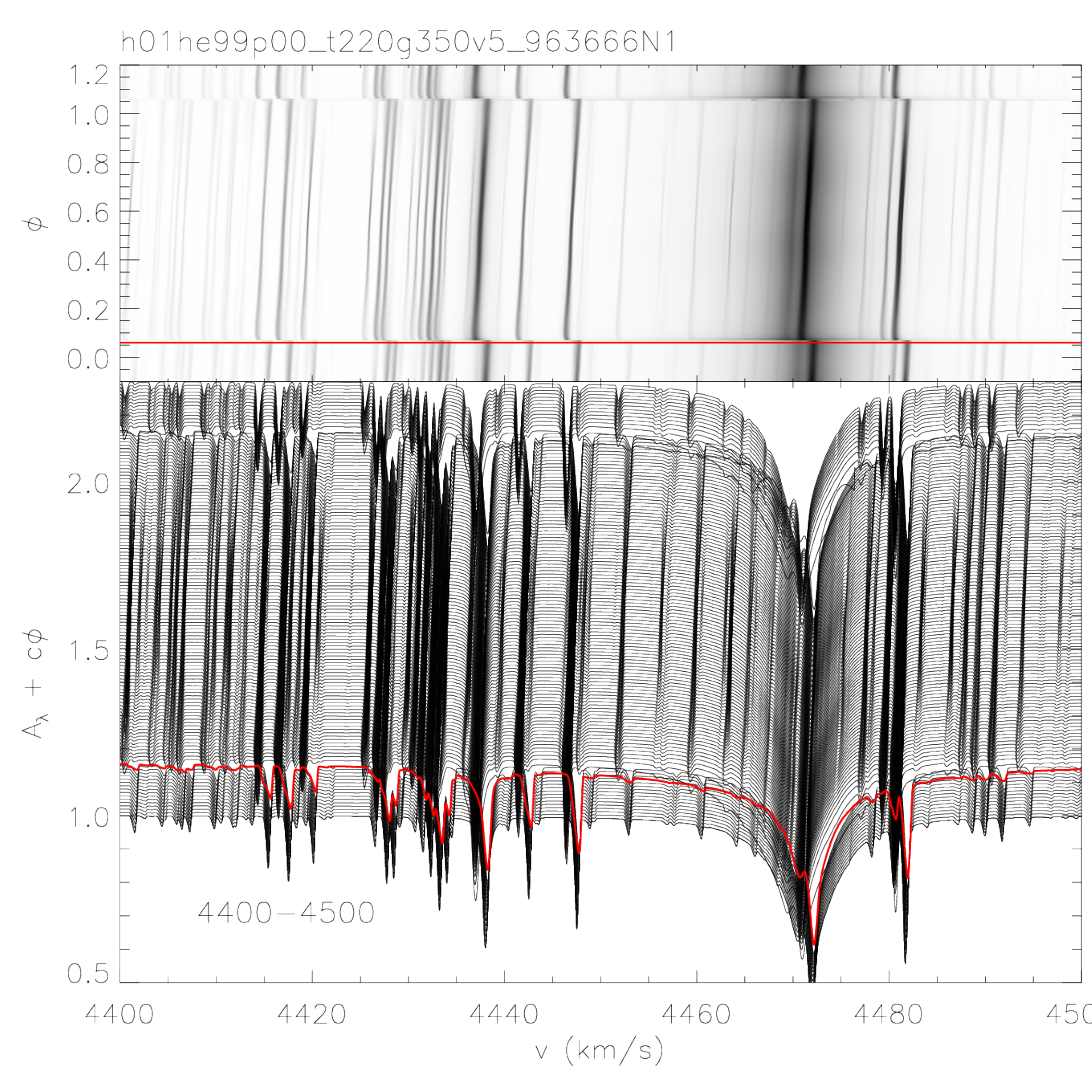}\\
\includegraphics[width=88mm,angle=0,clip=true]{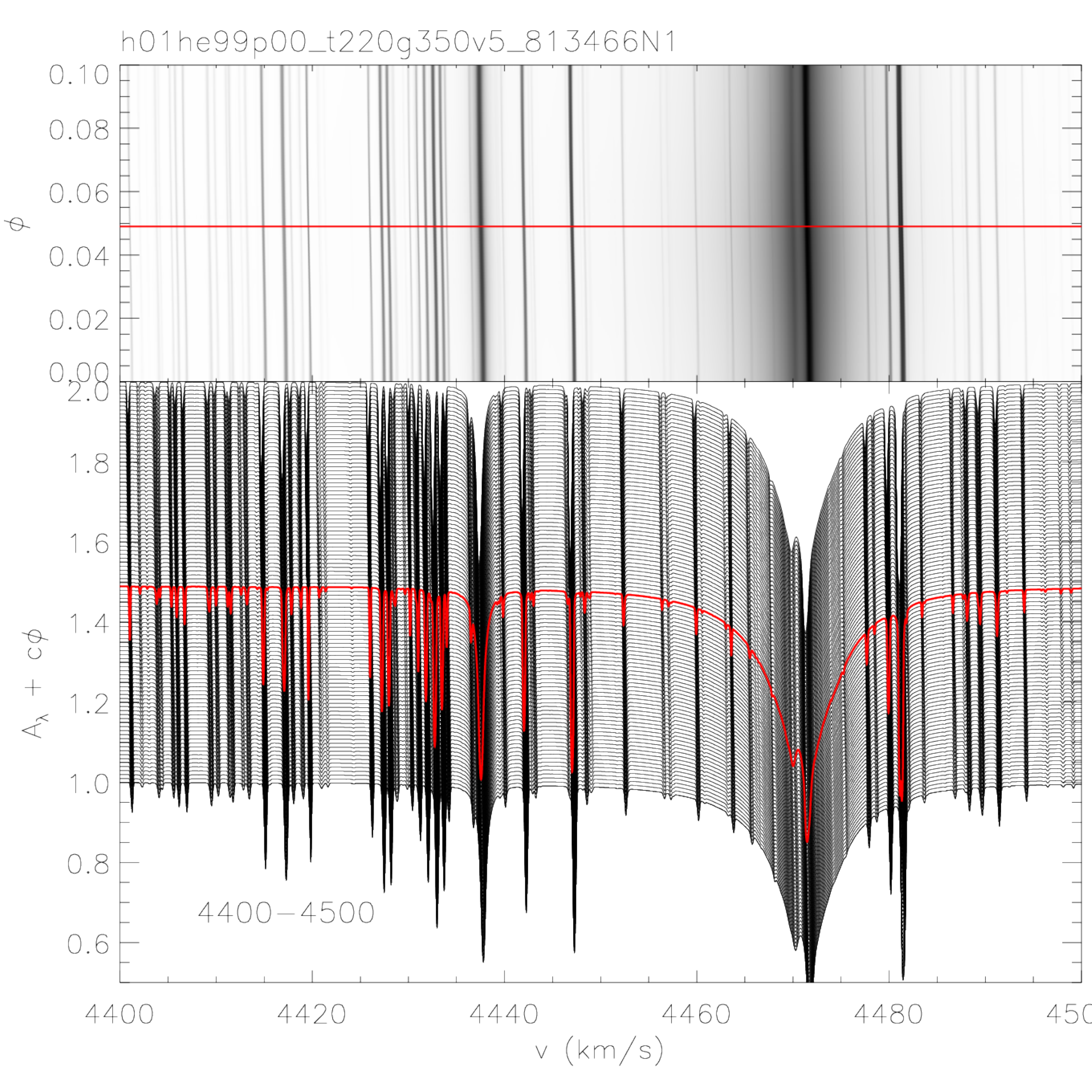}
\includegraphics[width=88mm,angle=0,clip=true]{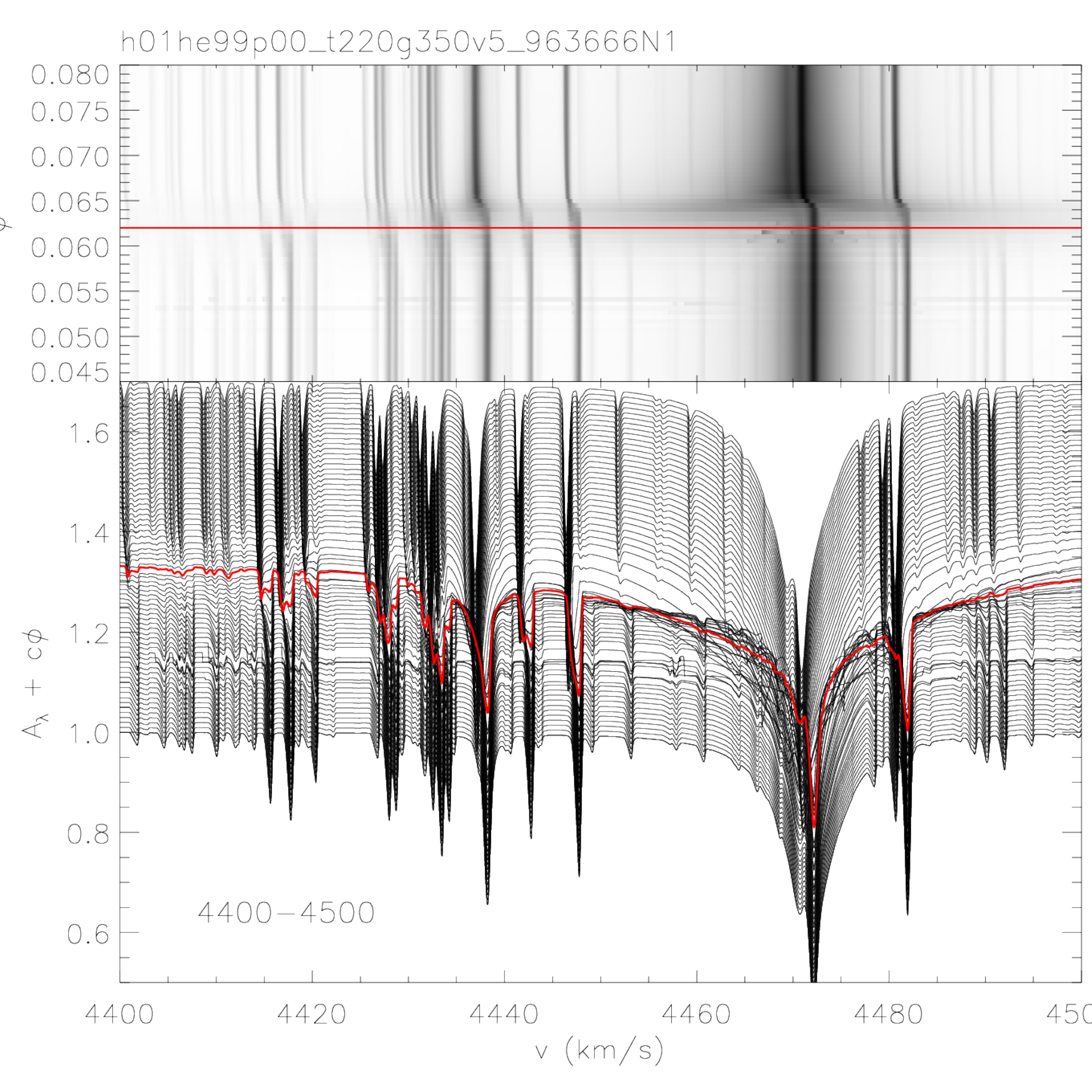}
\end{center}
\caption{As Fig.~\ref{f:shock_comp} for an extended part of the spectrum (4400 -- 4500\AA) obtained from two pulsation models. The region includes two strong He{\sc i} lines (4471\AA\ and 4347\AA) and a range of metal lines. As previously, the left-hand panels show the un-shocked pulsation model 814366N1 and the right-hand panels show the shocked model (963666N1). The top pair of panels show the entire pulsation cycle ($-0.1 < \phi < 1.2$) covering the shock phase twice over. The bottom pair of panels show the phase around minimum radius (shown in red in all panels). Note the variety of behaviours shown by different lines during the shock phase (lower right panel).   
}
\label{f:shock_synth}
\end{figure*}

\begin{figure*}
\begin{center}
\includegraphics[width=88mm,angle=0,clip=true]{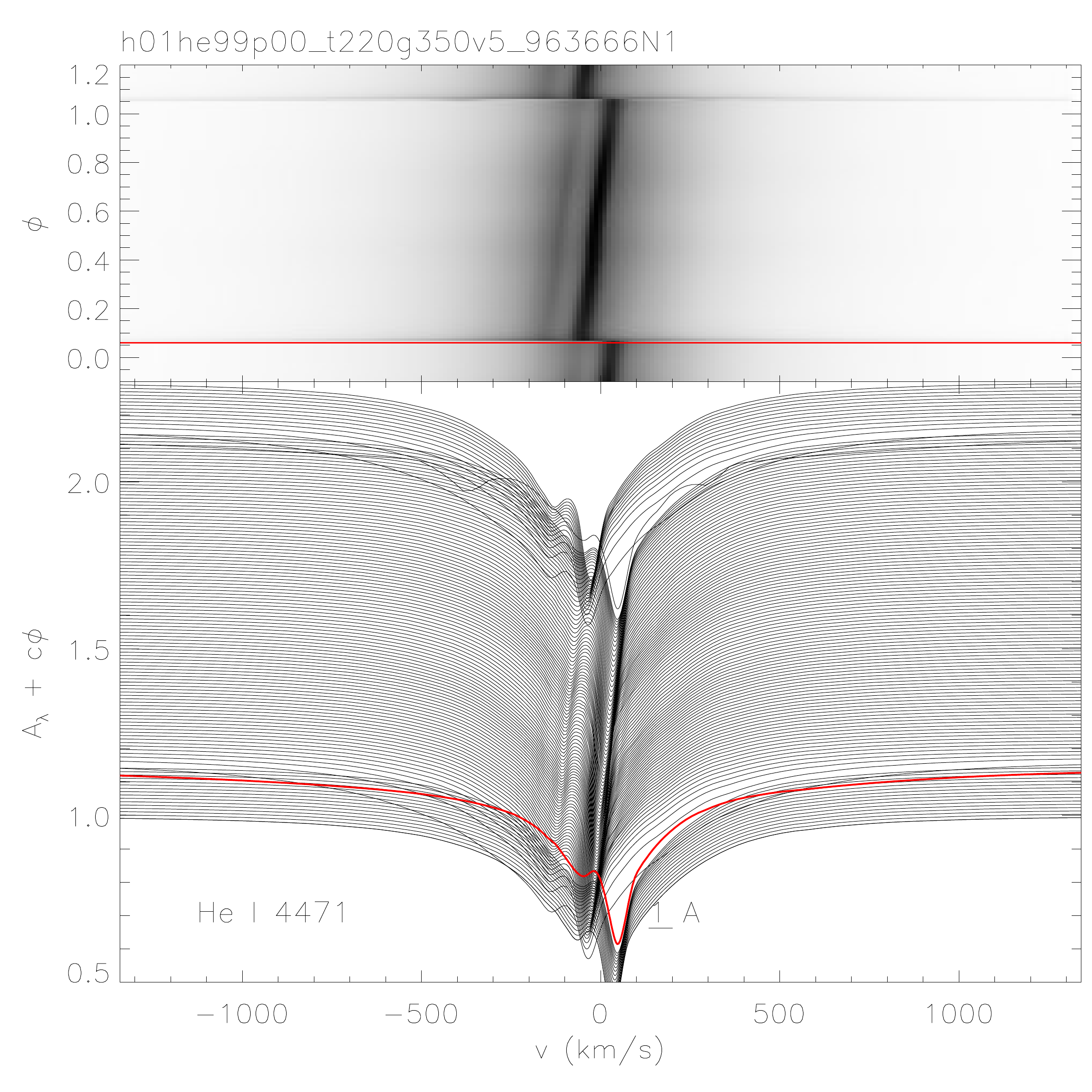}
\includegraphics[width=88mm,angle=0,clip=true]{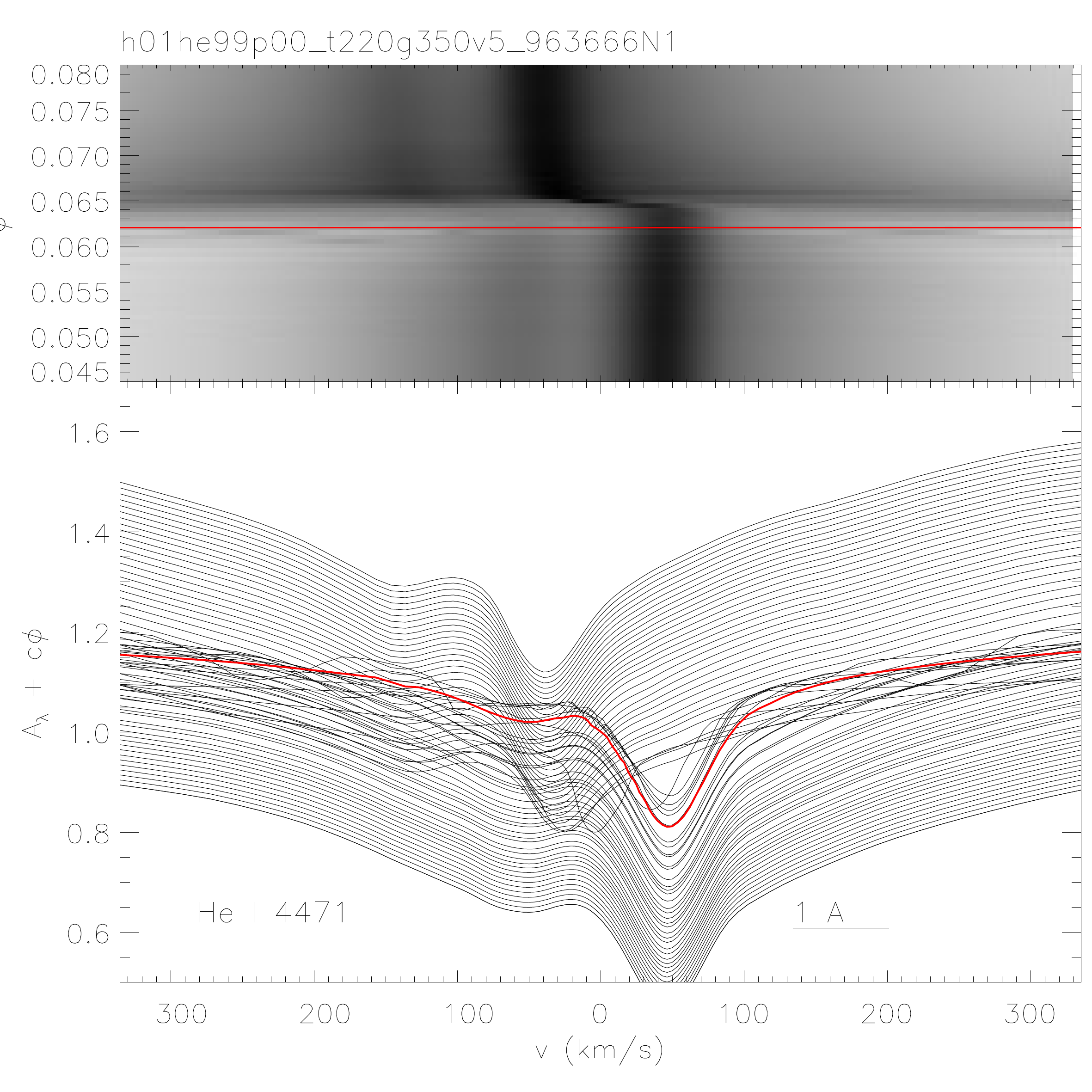}
\includegraphics[width=88mm,angle=0,clip=true]{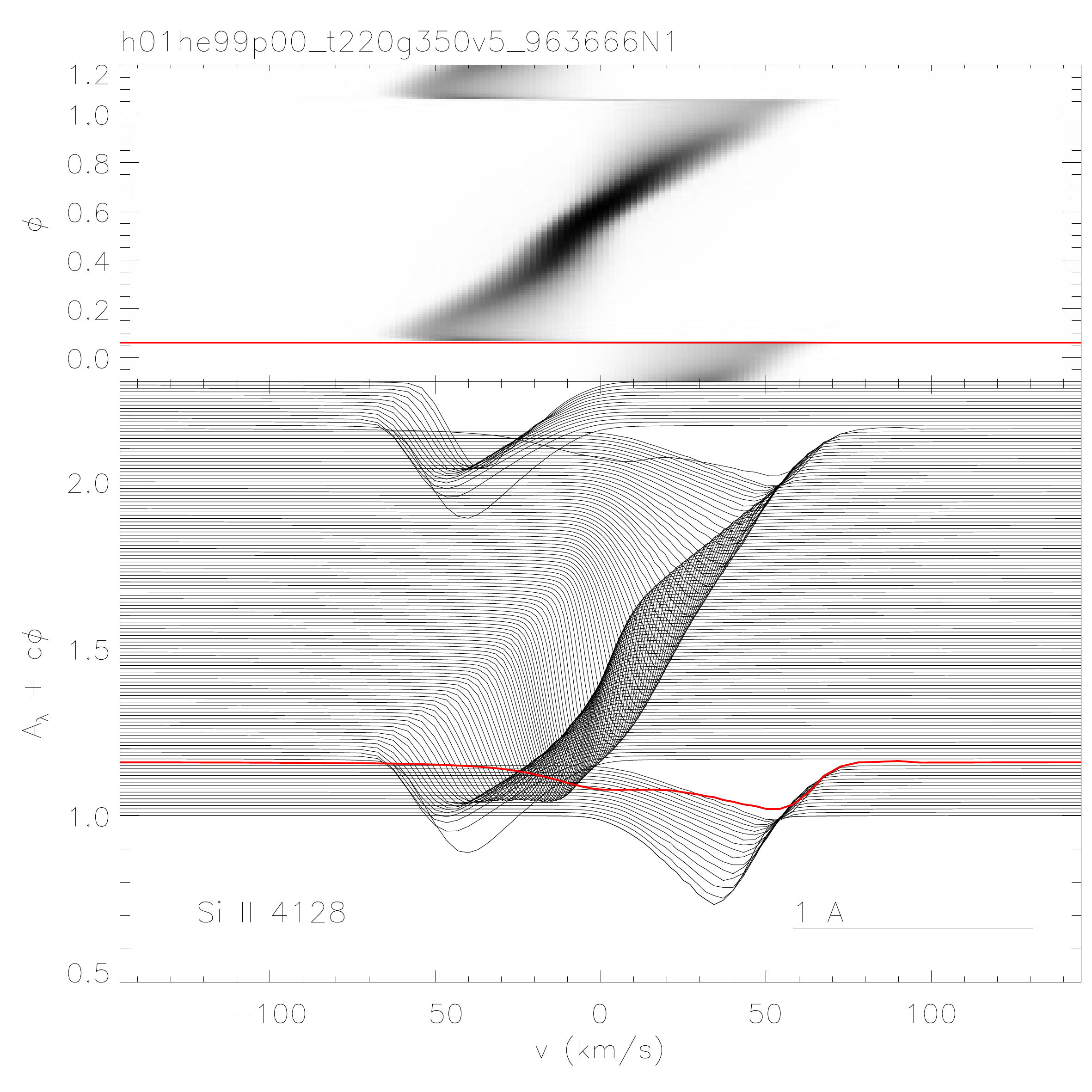}
\includegraphics[width=88mm,angle=0,clip=true]{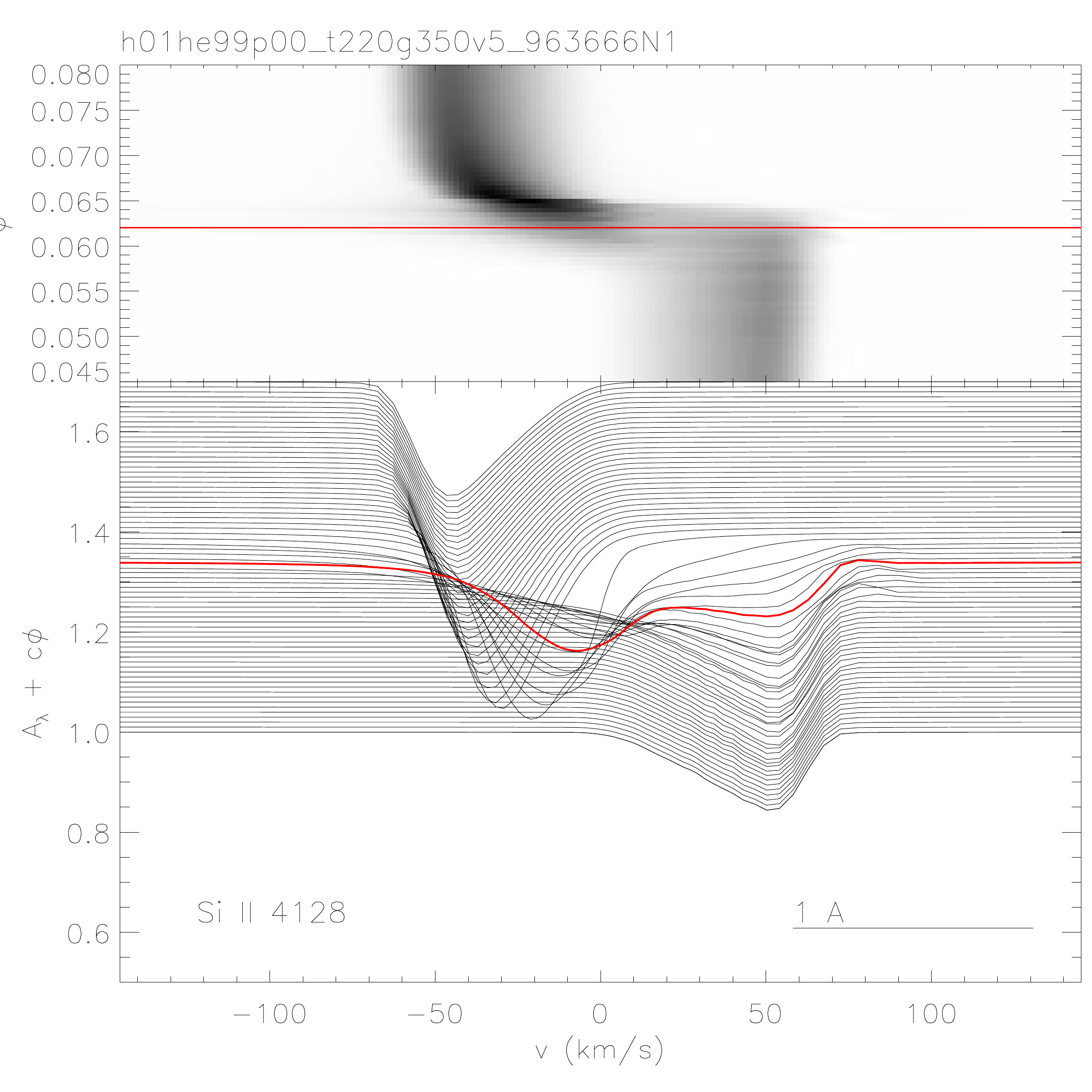}
\end{center}
\caption{As Fig.~\ref{f:shock_comp} showing detail of two contrasting lines, He{\sc i} 4471\AA\ (top) and Si{\sc ii} 4128\AA (bottom). The left-hand panels show the entire pulsation cycle ($-0.1 < \phi < 1.2$) covering the shock phase twice over. The right-hand panels show the phase around minimum radius (shown in red in all panels). 
}
\label{f:shock_detail}
\end{figure*}

\begin{table*}
    \centering
    \begin{tabular}{l ccc ccc}
    \hline
        Label &  $M / \Msolar$ & $\log L / \Lsolar$  & $\log \Teff / {\rm K}$ & $\sg / {\rm cm\,s^{-2}}$ & $\nH$ & Mix \\
        \hline
    \multicolumn{7}{l}{{\sc sterne} atmosphere models} \\
     h01he99p00\_t220g300  &  --   &  --  & 4.34 &  3.0 & 0.01 & p00 \\
     h01he99p00\_t220g350  &  --   &  --  & 4.34 &  3.5 & 0.01 & p00 \\[2mm]
    \multicolumn{5}{l}{{\sc nlpuls} pulsation models \citep{jeffery22a}} \\
    813466N1 & 0.66 & 2.81 & 4.34  & 3.78 & 0.01 & N1 \\
    913666N1 & 0.66 & 2.91 & 4.36  & 3.75 & 0.01 & N1 \\
    963666N1 & 0.66 & 2.96 & 4.36  & 3.69 & 0.01 & N1 \\
    064666N1 & 0.66 & 3.06 & 4.36  & 3.60 & 0.01 & N1 \\
    \hline
    \end{tabular}
    \caption{Models used as input for {\sc spec\_puls} test calculations. 'p00' implies a solar metal mixture. Mixture 'N1' is defined in \citet{jeffery22a} as $X=0.00125, Z=0.0159, Y\equiv1-X-Z$, and is based on \citet{przybilla05} (hydrogen) and \citet{jeffery01b} (metals).  }
    \label{t:input}
\end{table*}

\section{Evolution of a shock wave}
\label{s:shock}

An evaluation of {\sc spec\_puls} has been carried out using {\sc sterne} atmosphere and {\sc nlpuls} pulsation models (Table\,\ref{t:input}).
{\sc nlpuls} models are computed with a variable time-step which can be as short as 0.01\,s or as long as several seconds, depending on conditions. Each model sequence usually covers from 2 to 4 complete pulsation cycles. 
We define approximately 100 equally spaced phase points covering a region of interest, and then seek the timestamps corresponding most closely to those phases.
The pulsation model is then mapped onto the equilibrium model atmosphere grid points as defined in \S\,\ref{s:puls}. 
In practice, zoning of the pulsation model is limited to atmosphere optical depths $\log \tau > -3$, since the pulsation model does not currently extend above this region. 

Figs.\,\ref{f:rmin_813466} and \ref{f:rmin_963666} show the scaled pressure, scaled temperature and velocity structures of two of these models as functions of mass depth $m'$ and pulsation phase $\phi$. Mass depth is used in preference to optical depth, since the pulsation model is defined in terms of the former, and the relationship between the two is only defined for the equilibrium model. One may conveniently consider the mass range $-1.8<\log m' / {\rm g\,cm^{-2}} < 2<$ as corresponding to $-3<\log \tau<2.6$, with $\log m' \approx 0.7 \log \tau$ in the equilibrium model.

Only phases $0<\phi<0.1$ are shown, focused on minimum radius. 
Phase $\phi=0$ is defined to be the time of maximum light. 
Both models have pulsation periods $\approx0.11$\,d. 
Consequently phase intervals $\delta \phi = 0.1$ and 0.01 correspond, in this case, to time intervals of $\approx850$\,s and $\approx85$\,s respectively.  
We have adopted phase rather than absolute time as the time variable since the problem should be generalisable to stars with significantly shorter or longer periods. 

Fig.\,\ref{f:rmin_813466} shows a pulsation model with no shock.
The maximum over-pressure seen in the atmosphere barely exceeds a factor 5, and then only as the pressure wave reaches the outermost layers at minimum radius.
Heating reaches a maximum of 30 per cent above ambient at the base of the photosphere, again at minimum radius. 
The velocity structure of the atmosphere is uncomplicated; through minimum radius the atmosphere accelerates almost uniformly.

Fig.\,\ref{f:rmin_963666} shows a pulsation model with a shock-wave travelling outward through the photosphere. 
The over-pressure  in the atmosphere increases rapidly as the shock-wave travels into cooler less dense material with a lower sound speed, reaching a factor exceeding 80 in this case. 
Heating reaches a maximum of 80 per cent above the equilibrium temperature ($T_0$) at the base of the photosphere, again at minimum radius, with substantial heating both above and below this layer and especially along the shock front. 
The velocity structure of the atmosphere is characterised by a step-function which propagates outwards. Inward-falling material above (in front of) the shock decelerates sharply into a small stand-still region before being accelerated more smoothly once the shock has passed. 

\begin{figure*}
\begin{center}
\includegraphics[width=88mm,angle=0,clip=true]{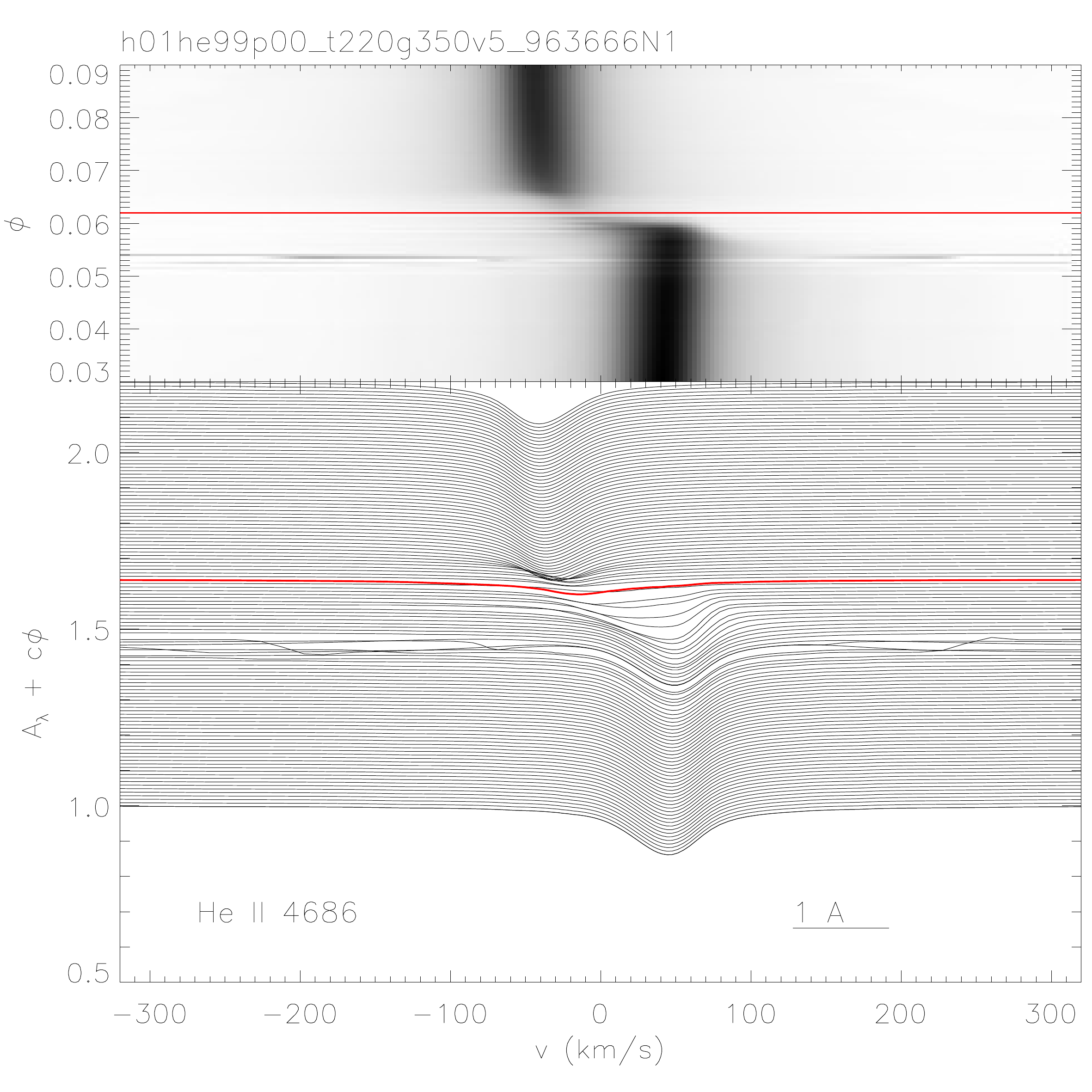}
\includegraphics[width=88mm,angle=0,clip=true]{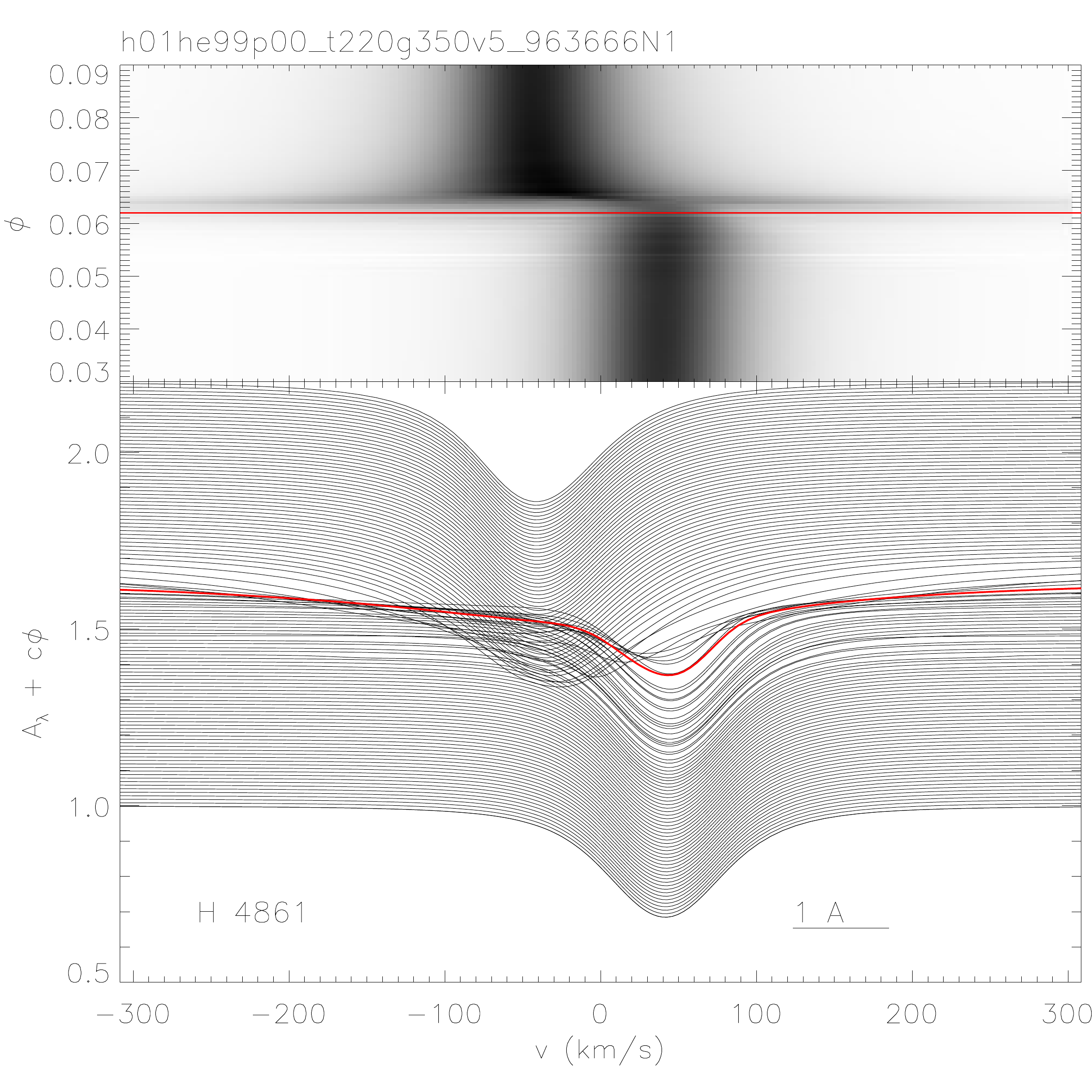}\\
\includegraphics[width=88mm,angle=0,clip=true]{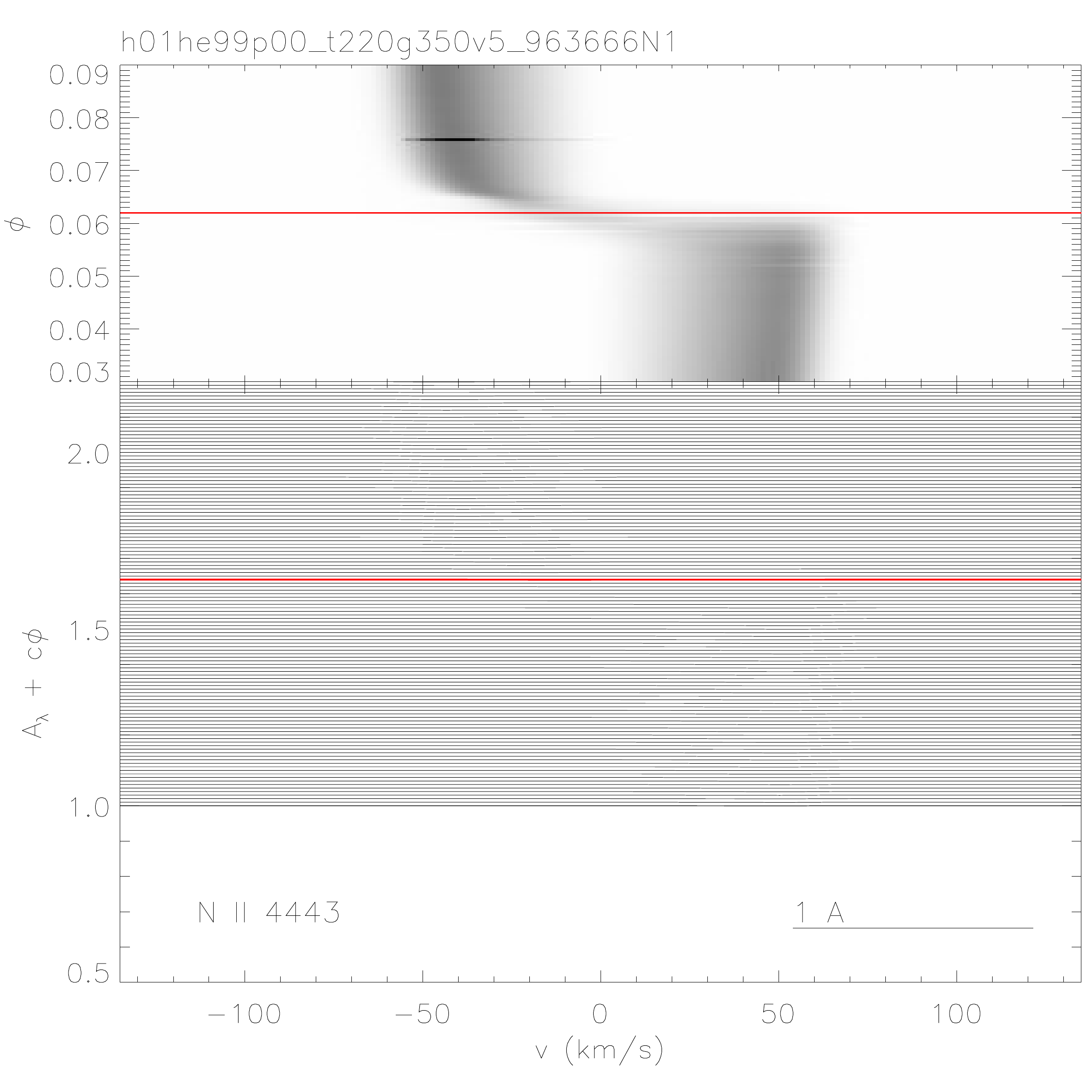}
\includegraphics[width=88mm,angle=0,clip=true]{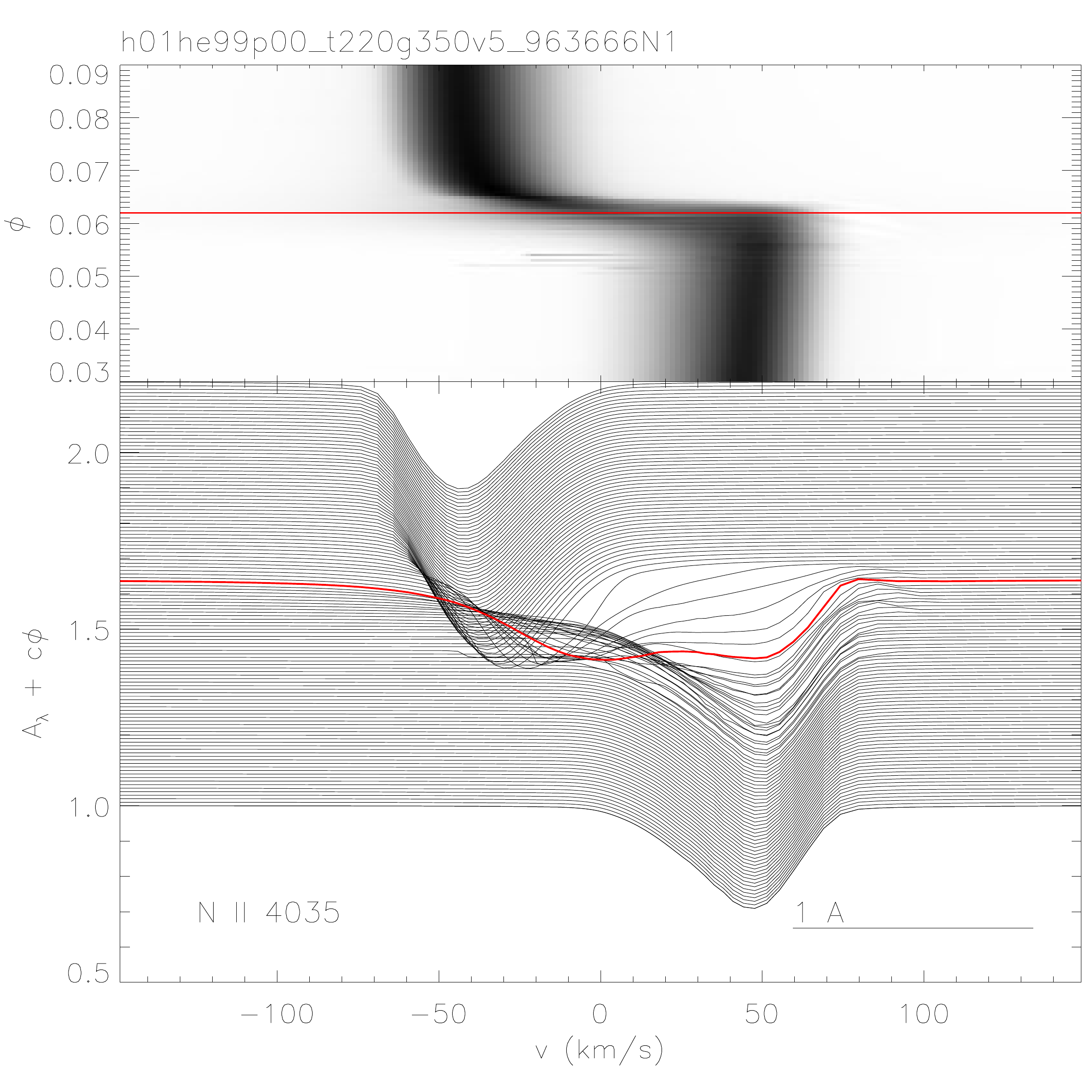}
\end{center}
\caption{As Fig.~\ref{f:rmin_963666} for 4 lines showing a variety of behaviours discussed in the text. The velocity scales for
He{\sc ii}4686\AA\ and H$\beta$ have been compressed in order to include more information from the line wings. With an equivalent width $<0.3$m\AA, N{\sc ii}4435\AA\  is invisible at the scale shown but is included for comparison with the much stronger N{\sc ii}4035\AA\ line; the lower-level excitation potentials of both lines are comparable. The grey-scale image shows how the shock reaches the weaker and deeper line sooner.  
}\label{f:shock_metal1}
\end{figure*}
\begin{figure*}
\begin{center}
\includegraphics[width=88mm,angle=0,clip=true]{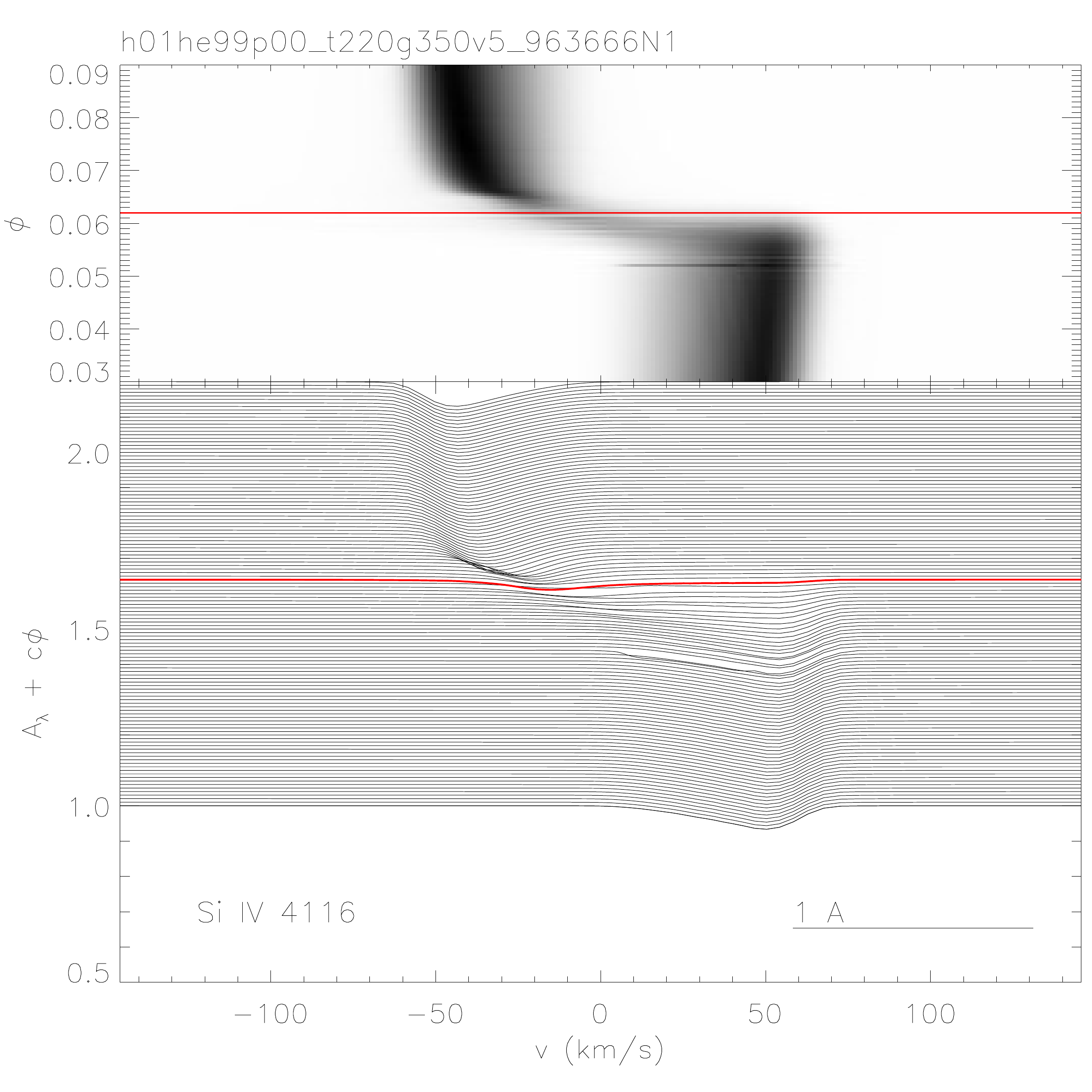}
\includegraphics[width=88mm,angle=0,clip=true]{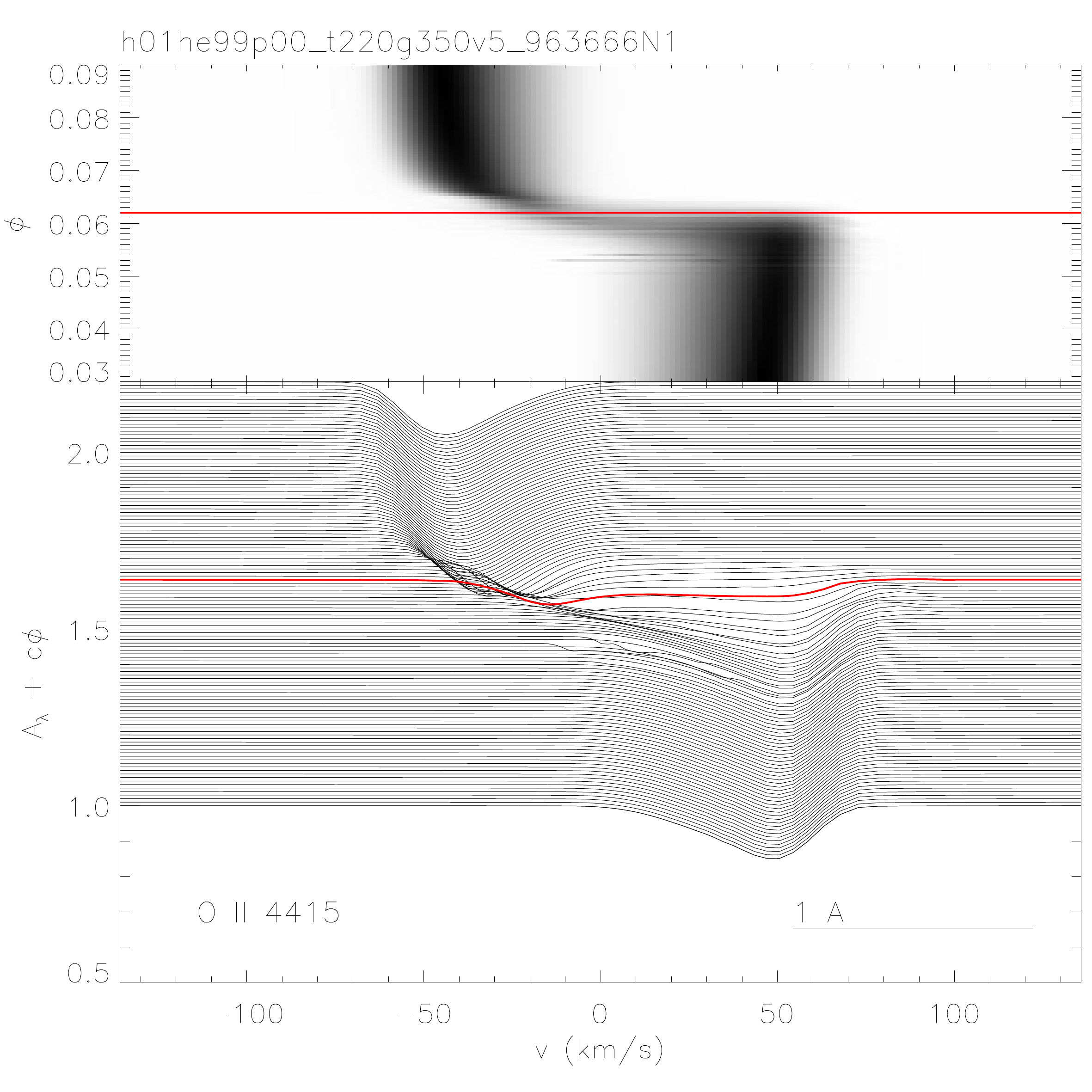}\\
\includegraphics[width=88mm,angle=0,clip=true]{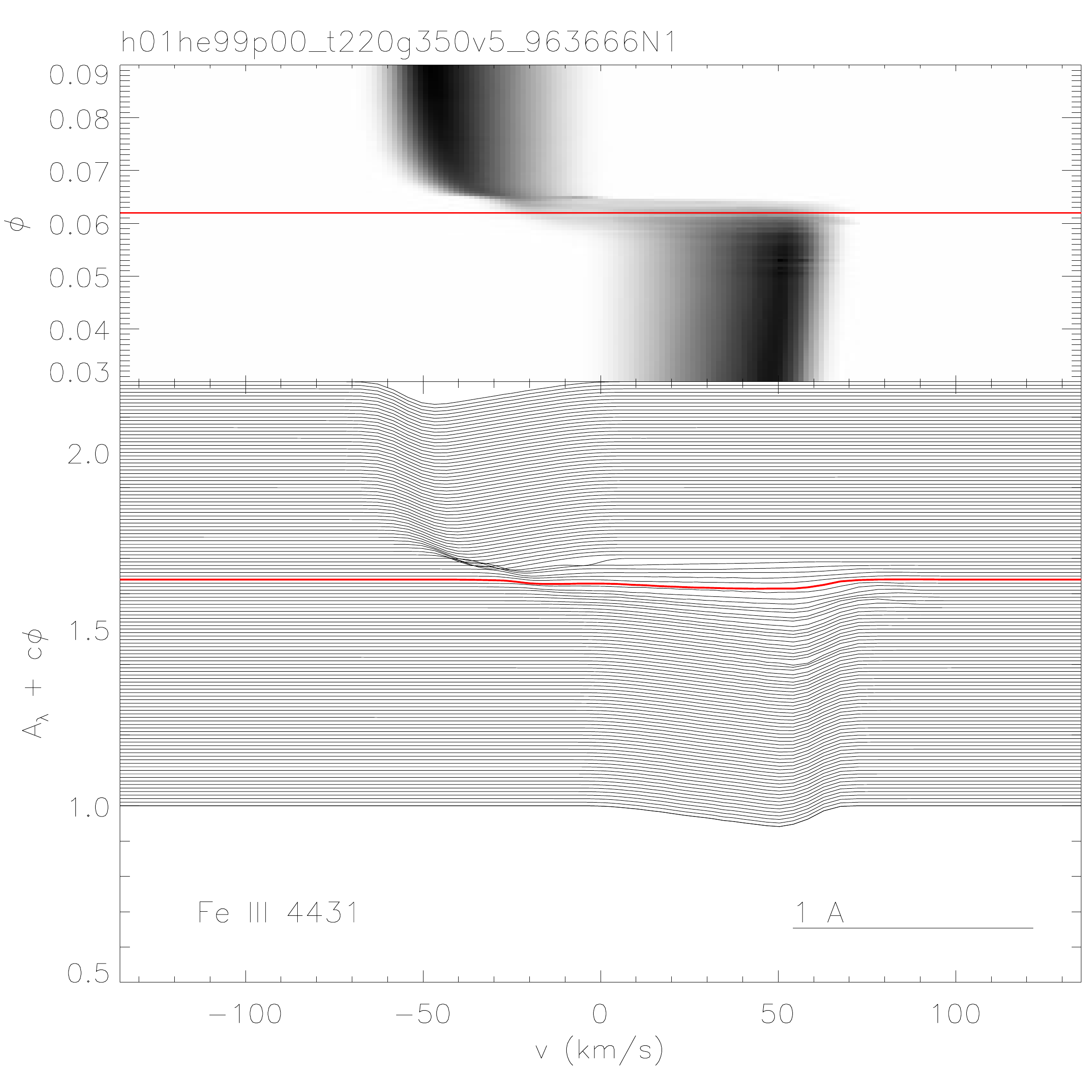}
\includegraphics[width=88mm,angle=0,clip=true]{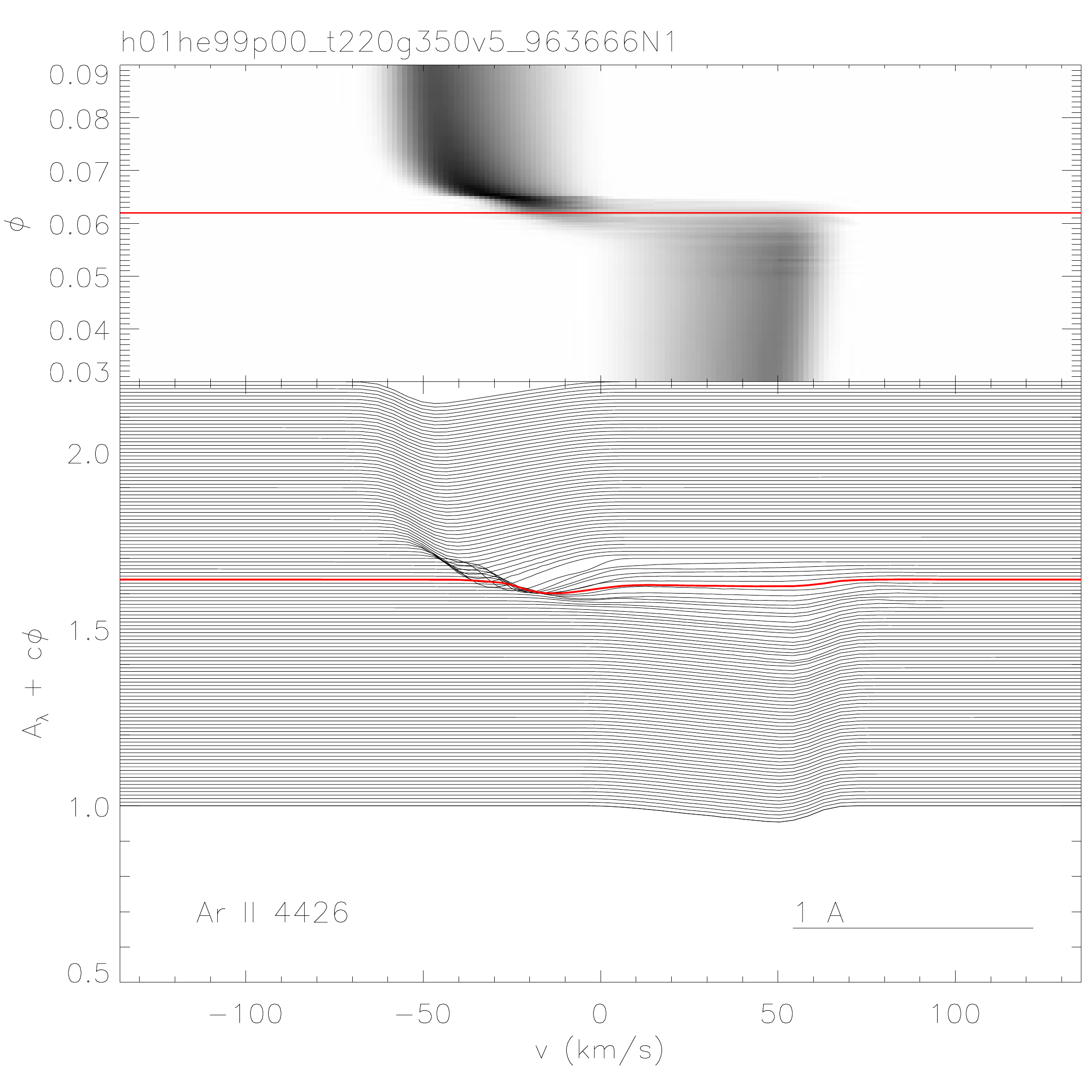}
\end{center}
\caption{As Fig.~\ref{f:shock_metal1} for 4 additional lines. }
\label{f:shock_metal2}
\end{figure*}

\section{Spectral synthesis}
\label{s:examples}

Fig.\,\ref{f:shock_comp} shows the predicted behaviour of an isolated strong line (Si{\sc iii}4553\AA) through radius minimum as computed from the two models described in the previous section. 
In the un-shocked model, the absorption line makes a smooth transition from maximum in-fall velocity (red-shift) to maximum expansion velocity (blue-shift). Since the  amplitude of the pulsation is not large (some 30 -- 40 km\,s$^{-1}$), the line is never strongly asymmetric.
In the shocked case, the line broadens strongly around minimum radius, with some hint of dividing into two components. The bluer component appears at approximately rest velocity before accelerating toward maximum expansion velocity, which is a negative wavelength shift in the observer's frame. The red-shifted component disappears at the same time.  

The response of an absorption line to variation in pressure, temperature and velocity varies  from line to line, depending on the effective temperature of the star, the ion in question, and its atomic properties. The cores of strong lines are formed higher in the atmosphere than those of weak lines. Response to temperature depends on the local temperature relative to the ionization 
energy for that ion. Response to pressure depends on the strength of the line and its sensitivity to the local electric field (damping). Fig.~\ref{f:shock_synth} shows 100\AA\ of spectrum in the blue-optical for both pulsation models (813466N1 and 963666N1). 
The top panels show the full pulsation cycle,  the bottom panels show the same models expanded around minimum radius. Several line behaviours are evident. 

In the un-shocked model (8134666N1) there is a slight broadening of the He{\sc i}4471\AA\ line near minimum radius, which occurs at $\phi = 0.049$. This is primarily due to a local increase in pressure and/or effective surface gravity (the two are  connected) whilst the surface acceleration ($\ddot{r}$) is positive. Otherwise most lines behave approximately as  Si{\sc iii}4553\AA\ line in Fig.~\ref{f:shock_comp}. 

In the shocked model (963666N1) the broadening of the He{\sc i}4471\AA\  is dramatic. Since the line wings are primarily formed in deep layers, the line starts to broaden some time before minimum radius ($\phi = 0.062$) as these layers are compressed by the outward moving pressure wave. 
The broadening extends up to $\pm 20$\AA\ on either side of the line centre. 
Remarkably, the far wings reach maximum strength earlier in the cycle than the near wings, at about the same time as the line core starts to shift blue-ward. 
As the core shifts, so also does the location of maximum strength in the wings, giving the latter a bow-shaped appearance in Fig.~\ref{f:shock_detail} (upper right). 

The wings track the shock through the line forming region. 
The phase of maximum broadening corresponds to that of maximum line shift (acceleration) and occurs $\approx0.003$ cycles {\it after} minimum radius.   
Fig.~\ref{f:shock_detail} (upper panels) provides more detail by treating  He{\sc i}4471\AA\ as an isolated blend.  Other strong lines with damping wings showing a similar effect can be identified from Fig.~\ref{f:shock_synth}.

In the grey-scale plot of Fig.~\ref{f:shock_synth}, the cores of well isolated weak lines show doubling up to 0.004 cycles before minimum radius, and  respond earlier than the cores of strong lines, as should be anticipated. 
As in Si{\sc iii}4153\AA, the component associated with in-falling material above (in front of) the shock remains at high red-shift, weakening gradually as the shock progresses through the photosphere. 
The component associated with accelerated material below (behind) the shock appears close to rest velocity and then becomes increasingly blue-shifted as the shock advances in front of it. 

Fig.~\ref{f:shock_detail} (lower panels) shows this in more detail by treating  Si{\sc ii}4128\AA\ as an isolated line. 
Several things should be noted from these two panels.
First, the temperature of this star is such that the atmospheric abundance of the Si$^+$ ion drops with increasing temperature.
Second, combined with line narrowing, Si{\sc ii}4128\AA\ consequently appears strongest (deepest) at maximum radius. 
Third, compression increases the ion number density, making the line appear  briefly deeper as the shock passes. 
Fourth, this and similar lines could suggest a brief stand-still phase for this model. 
This has not yet been observed in V652\,Her.    
Lines due to other ions with similar temperature characteristics (e.g. Mg{\sc ii}, Al{\sc ii}, P{\sc ii}, S{\sc ii}, Ar{\sc ii}) show similar behaviour.  

Figs.~\ref{f:shock_metal1} and \ref{f:shock_metal2} show the behaviour of eight additional lines during shock passage.
These have been chosen to be representative of lines formed at different depths, but also having significant differences and similarities. 

For He{\sc ii}4686\AA, the transition from red- to blue-shift commences 0.002 cycles before minimum radius and is accompanied by a substantial weakening of the line (Fig.~\ref{f:shock_metal1}) through minimum radius. 
In contrast H$\beta$ behaves like the He{\sc i} lines; it commences blue-shift late and is accompanied by strong wing-strengthening with a bow-shaped signature.  

The comparison of a very weak and a very strong N{\sc ii} line is instructive. Blue-shift commences earlier in the weaker (deeper) line. Doubling is evident in the stronger (shallower) line, but is difficult to discern in the weak line. 

Blue-shift also occurs early in the high-ionization line Si{\sc iv}4416 (Fig.~\ref{f:shock_metal2}), which must necessarily form deeper in the atmosphere than lines from lower ionization species (cf. Si{\sc iii}4553\AA\ Fig.~\ref{f:shock_comp} and Si{\sc ii}4128\AA\ Fig.~\ref{f:shock_detail}). O{\sc ii}4415\AA\ and Fe{\sc iii}4431\AA\ represent a pair of lines with very similar evolution, but significantly different phases. Ar{\sc ii}4426\AA\ resembles Si{\sc ii}4128\AA\ in some respects, and is the opposite of He{\sc ii}4686\AA\, since it strengthens momentarily just after shock passage, although the shock reaches this line  long before it reaches Si{\sc ii}4128\AA. 

We conclude that the response of each spectral line to the passage of a shock wave depends on several properties including the line depth, the ionization degree and temperature response of the parent ion, and the response of the line to pressure broadening. 

\begin{figure*}
\begin{center}
\includegraphics[width=175mm,angle=0,clip=true]{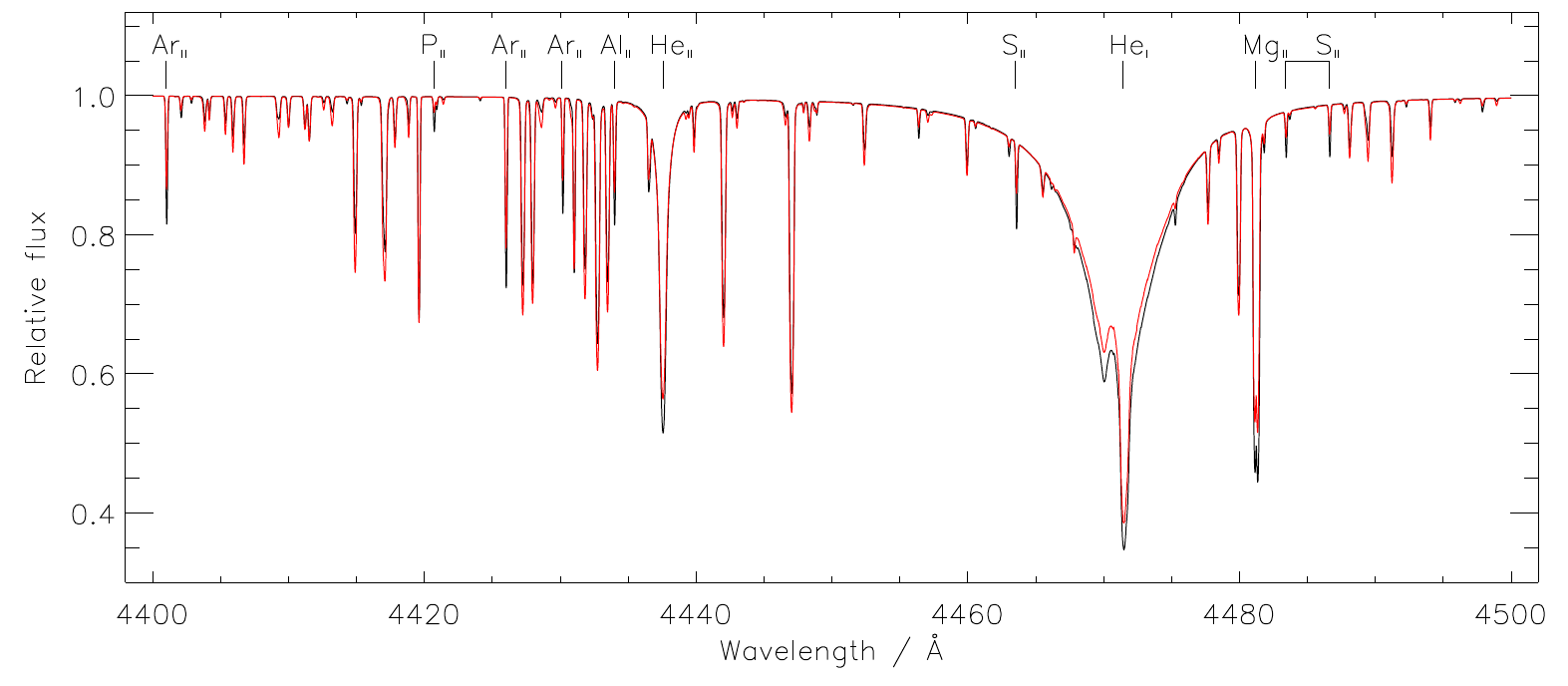}
\end{center}
\caption{Part of the spectrum obtained from the reference model atmosphere (h01he99p00\_t220g350v5) combined with the pulsation model (963666N1) at maximum radius (black) and assuming that the same model atmosphere is in complete hydrostatic equilibrium (red). No additional broadening has been applied to either model. Lines which are deeper in the pulsation model are labelled with their ion. Other (strong) lines may be identified with reference to \citet[][Fig. B1: part c -- additional material online]{jeffery15b}. 
}
\label{f:rmax_static}
\end{figure*}

Fig.~\ref{f:rmax_static} compares a part of the spectrum from the pulsating atmosphere at maximum radius (black) with the spectrum obtained from the same atmosphere (h01he99p00\_t220g350v5) in hydrostatic equilibrium (red). 

In this figure, with the exception of He{\sc i}, lines may be divided into (a) those arising from ions with low ionization potentials (e.g. Mg{\sc ii}, Al{\sc ii}, P{\sc ii}, S{\sc ii}, Ar{\sc ii}) in which the hydrostatic line is weaker than the pulsating line, and (b) those arising from ions with higher ionization potentials (e.g. N{\sc ii}, O{\sc ii}, Ne{\sc ii}, Al{\sc iii}, Si{\sc iii}, S{\sc iii}, Fe{\sc iii}) \citep{jeffery15b} in which the hydrostatic line is stronger than the pulsating line.  
Both are a consequence of the stellar surface  at maximum radius being cooler than the equilibrium value. 
If one were to use the spectrum at maximum radius to infer the mean effective temperature by solving for the ionization equilibrium, one would underestimate the equilibrium value by $\approx 2000$\,K.  

By the same argument, a lower effective temperature should result in stronger He{\sc i} lines.  In the example shown  (Fig.~\ref{f:rmax_static}),  the wings of the Stark-broadened He{\sc i} lines are not noticeably different between the dynamical model and that in hydrostatic equilibrium. 
In an example using the same pulsation model (963666N1) and a model atmosphere having $\log g / {\rm cm\,s^{-2}} = 3.00$ (h01he99p00\_t220g300v5), the Stark-broadened He{\sc i} line wings are weaker in the dynamical model than in hydrostatic equilibrium. 
This may be understood by the considering the surface deceleration at maximum radius. 
This reduces the effective surface gravity ($g_{\rm eff} = g + \ddot{r}: \ddot{r}<0$) and hence the local electron density, thereby reducing the pressure broadening and weakening the line wings.   
To understand the difference in impact between the two models, consider the relative contribution of the surface deceleration in each case.
Following \citet{jeffery15b}, the surface deceleration of V652\,Her at maximum radius $\ddot{r} \approx -1200\,{\rm cm\,s^{-2}}$. 
This is comparable with the equilibrium gravity at $\log g / {\rm cm\,s^{-2}} = 3.00$, but approximately one third that at $\log g / {\rm cm\,s^{-2}} = 3.50$. For a given surface deceleration, the relative contribution is thus greater at low surface gravity. 
A measurement based on fitting the He{\sc i} line wings at maximum radius may be too low by $\approx 0.25$ dex at $\log g / {\rm cm\,s^{-2}} = 3.5$, increasing to $\approx 0.5$ dex at $\log g / {\rm cm\,s^{-2}} = 3.0$.  

In the course of preparing this paper, other combinations of pulsation and atmosphere models were considered. 

Pulsation model 913666N1 (Table\,\ref{t:input}) was selected as an example of a borderline shock. Although the radial velocity shows a very steep acceleration phase, no shock forms and the line behaviour at minimum radius resembles that of model 8134666N1, albeit with acceleration over a smaller phase interval ($\approx 0.025$ cycles compared with $\approx 0.06$ cycles). 

Pulsation model 064066N1 (Table\,\ref{t:input}) was selected as an example of a more extreme shock, with a radial velocity amplitude $\approx140\kmsec$ (compared with $\approx120\kmsec$ for 963666N1). The velocity amplitude excepted, the individual line diagnostics are similar for the two cases. 

A series of calculations was run for the the model atmosphere h01he99p00\_t220g300v5. The lower surface gravity $\log g / {\rm cm\,s^{-2}} = 3.0$ rather than $3.5$  means that all lines and especially the Stark-broadened lines are  narrower and deeper, giving better resolution in the line diagnostic plots. 
There is slightly less contrast between the in and out of shock line depths in the case of He{\sc 4686}\AA\ and Ar{\sc ii}4426\AA, but the behaviours are essentially the same. 

\section{Discussion}
\label{s:discussion}

Several questions arise.  For discussion of these and unless otherwise indicated, "model" refers to the theoretical line profiles and spectral syntheses computed for this paper, and not to the underlying pulsation or atmosphere models used as inputs. 

\paragraph*{How do the models compare with models for other classes of pulsating star?}

Theoretical comparators based on non-linear pulsation models for radially pulsating stars are sparse. 
\citet{sasselov92} show model profiles for a Mg{\sc ii} line around the pulsation cycle of the classical Cepheid $\zeta$ Geminorum; the sequence around minimum radius is sparsely sampled but shows broadening and probably two components. 
\citet{fokin04} show model profiles for a Si{\sc iii} line around the cycle of the $\beta$ Cepheid BW\,Vulpeculae. Again sparsely sampled, their model (their Fig.\,8) strongly resembles our own  during the shock phase (Fig.\,\ref{f:shock_comp}); a component appears close to stand-still velocity and then migrates to the blue just as the red-shifted component disappears. 
\citet{chadid08} show the profile of an Fe{\sc ii} line observed through minimum radius of the RR\,Lyrae star S\,Arae (their Fig. 7). The line shows precisely the same behaviour just described for our models and those of \cite{fokin04}. 
While it would be valuable to find additional comparators, it is  encouraging that the line behaviour seen in our models has been identified in previous theoretical and observational investigations.

Whilst the  {\sc spec\_puls} method is comparable to that of \citet{fokin90}, it differs by separating the  calculation of the pulsating interior from that of the photosphere structure and from the formal solution of the emergent spectrum. Whilst a fully self-consistent scheme is desirable, {\sc spec\_puls} was suggested by our familiarity with well-tested codes. Separating the code elements substantially reduces the computational cost, particularly for the hydro calculation, and also offers flexibility to experiment with different methods or physics within each component. 

\paragraph*{How do the models compare with observations of V652 Her and BX Cir?}

Subaru Telescope high-resolution spectroscopy of V652\,Her with exposure times of 120\,s was presented by \citet{jeffery15b}.
Using groups of lines formed at different depths and using the phase of maximum acceleration in each group, the latter (their Fig. 12) mapped the passage of the shock through the photosphere over an interval of $< 0.02$ cycles (or $<180$\,s). 
Our shocked models show a shock transit time $\approx 0.015$ cycles (our Fig.\ref{f:rmin_963666}).
Line doubling was not observed directly, but equally there was no smooth migration of the red component to the blue through minimum radius (their Figs. 14 - 16). 
Higher time resolution in the observations would help.  
There is evidence of different behaviours in different lines, but this also needs further examination.  

SALT high-resolution spectroscopy of BX\,Cir has been reported by \citet{martin19.phd}. 
No evidence was found for different behaviours in lines of different strengths. 
Since the radial-velocity curve does not show a sharp transition at minimum radius \citep{kilkenny99,woolf02}, it appears that no shock is formed and hence we expect the photosphere to accelerate as a single entity through minimum radius (Figs.\,\ref{f:rmin_813466}, \ref{f:shock_comp}:left and \ref{f:shock_synth}:left).

\paragraph*{What have we learnt about shock diagnostics for pulsating stars?}

Line doubling is well-established as a likely indicator of an atmospheric shock-wave travelling through the atmosphere \citep{schwarzschild52} and is indicated by our models. Phase-dependence of shock passage with line depth has been observed and used for tomography in long-period variables \citep{alvarez00}.

We have shown that substantial compression of material in the shock should be observable in the He{\sc i} line wings, and also that the position of maximum broadening is phase dependent and tracks the shock passage.  

The strength of the blue component immediately after formation varies from ion to ion as well as with line depth.  
This is best illustrated by comparing Ar{\sc ii}4426\AA\ and He{\sc ii}4686\AA\ (Figs.\,\ref{f:shock_metal1} and \ref{f:shock_metal2}).
The positions of both red and blue components are comparable with each other and with those in other lines. 
At maximum acceleration  of the blue component, Ar{\sc ii}4426\AA\ line becomes markedly deeper.
He{\sc ii}4686\AA\ becomes substantially weaker; in the model shown it almost disappears around minimum radius. 
Both phenomena suggest a sudden {\it cooling} of the material where the line is formed, but there is no indication of any cooling relative to ambient in the pulsation model (Fig.~\ref{f:rmin_963666}).
Rather, we surmise that non-adiabatic compression in the shock induces recombination via the electron-density term in the Saha equation, which here dominates the temperature term. 
The question deserves further exploration but is out of scope for this paper.

\paragraph*{What prospects are there for additional tests of the new models?}

The detection of line-to-line difference in the behaviour of line profiles though minimum radius, including both phase delay and strength variation, would provide a clear test of the pulsation models and additional insight into shock physics.
Tomography has the potential to resolve velocity, pressure (from Stark broadening) and temperature (from ionization balance) structure within the accelerating photosphere.  
Observations of the shock phase for V652\,Her are challenging; the shock phase lasts less than 200\,s and tests require high-resolution and high signal-to-noise observations.  
The Subaru and Space Telescope data already available have not been fully exploited; additional steps can be made to improve phase resolution and to explore more lines over a greater range of atmospheric depths. 
It is hoped that these can be realised in the near future. 

\paragraph*{How might the models be further improved?}

Given resource, a larger volume of parameter space should be explored, for both pulsation and atmosphere models. In particular, model atmospheres covering a wider range in temperature should be investigated, although the boundaries of the instability strips where radially pulsating stars are found should be observed. 

At present the pulsation model extends to optical depths $\log \tau > -3$. It would be valuable to extend the calculation to lower masses, in particular to remove the outer boundary as far as possible from the line-forming region. 

Rather than using the radial velocities directly from the pulsation model, the relative radius perturbations should be mapped onto the model atmosphere and converted to radial velocity thereafter. 
This will improve accuracy when using the models to extract physical information from observational data and so must be done in due course.
It does not affect conclusions from the model exploration exercise. 

With substantial compression and heating factors, the physics modules should be re-examined for treatment of  electron degeneracy, pressure ionization and departures from LTE. 
Since a shock traverses the atmosphere on a timescale of seconds, a time-dependent equation of state in which level populations are assumed not to be in either local thermodynamic or statistical equilibrium might ultimately be necessary,
 
Before that, in-homogeneous models should be constructed in order to investigate the r\^ole of chemical diffusion on the dynamics of pulsations in blue large-amplitude pulsators \citep{byrne20}.  

 A non-LTE treatment of line formation will have consequences, particularly for strong lines, as this is known to be important in V652\,Her \citep{przybilla05}. However it is impossible to predict how the behaviour of any given line through shock passage might be modified by such a treatment. One might look for potential emission due to a population inversion, or for an earlier appearance or a  delayed disappearance of line components. A non-LTE treatment for all lines would substantially increase computation times, since occupation numbers would have to be computed for every layer at every time step.

A more sophisticated treatment of shock passage using a 3D simulation, would be instructive. Shock-induced turbulence might be evident from additional line broadening, but further consequences are beyond the scope of this paper. 

\paragraph*{What are the limitations of {\sc spec\_puls} and can it be used to investigate other pulsating stars?} 

 The primary motivation for developing {\sc spec\_puls} was to construct dynamical models for the spectrum of the pulsating helium star V652\,Her \citep{jeffery15b}.
{\sc nlpuls} was originally written to investigate Type II Cepheids \citep{bridger83}. 
Based on a program for the atmospheres of central stars of planetary nebulae \citep{bohm65}, {\sc sterne} was developed for early-type helium stellar atmospheres \citep{hunger67,wolf73} but has been extensively used for hydrogen-rich stars \citep[e.g.][and following]{heber84} and modernised in other ways \citep[e.g.][]{behara06}.
 Indeed,  many  early model atmosphere would adopt chemical abundances relative to hydrogen and hence fail when abundance of the latter was set to zero. At least for hot stars, the physics treated in {\sc sterne} is equivalent to that used in {\sc atlas12} \citep{kurucz05}. 
{\sc spectrum} was developed for main-sequence B stars \citep{dufton72}, later assuming  {\sc atlas6} model atmospheres \citep{kurucz79} as input. 
It  was subsequently adapted for more general mixtures and model atmosphere inputs \citep{jeffery92}, and has been rebuilt in modular form so that different physics packages can be easily introduced. 
 Decades developing and working with {\sc nlpuls}, {\sc sterne} and {\sc spectrum} have made them the tools of choice for developing {\sc spec\_puls}. 

 {\sc spec\_puls} clearly has limitations: (i) none of the codes  makes any provision for molecules, (ii) {\sc nlpuls} uses the Courant condition, which limits mass-zone and time-step sizes to avoid numerical problems when shocks occur, (iii) {\sc spectrum} uses an LTE equation of state, so cannot help with models involving shock-induced emission lines (see above),  and (iv) 3D effects such as turbulence in the shock front or the consequences of rapid rotation cannot be modelled.  

Otherwise, {\sc spec\_puls} is easily capable of modelling radial-mode pulsations in any early-type star (e.g. $\beta$ Cepheids, blue large-amplitude pulsators, pulsating subdwarf B stars), as well as being applicable to RR\,Lyraes and classical Cepheids.  Other non-linear pulsation and/or stellar atmosphere models (e.g. {\sc atlas9, atlas12}) could be assimilated by providing suitable a interface.

\section{Conclusion}
\label{s:conclusion}

A suite of computer programs has been adapted for the computation of emergent spectra from  radially-pulsating stars. It operates by scaling the temperature and pressure perturbations from a non-linear pulsation model to an equilibrium model, by applying these scaling factors to an equilibrium model atmosphere at the same mass depths, and by incorporating the expansion velocities from the pulsation model. The codes are general.

Results for models applicable to two radially pulsating extreme helium stars, namely V652\,Her and BX\,Cir, show good qualitative agreement with the observations. 
Differences are demonstrated between models in which a shock wave develops in the atmosphere, and models where it does not.  

For shocked atmospheres, it is already known that shock progression is traced by different lines as a function of their strength, since the strength of a line is related to its depth of formation. 
Our models show that progression is also traced by lines as a function of their parent ion properties and may thus provide additional information about physical conditions in the shock.  
The behaviour of the different ions immediately post-shock points to a drop in the ionization temperature. 
The progression of the shock is also tracked by the wings of Stark-broadened absorption lines. 
Together, these models suggest that tomographic mapping could track the velocity, pressure and temperature structure of a pulsation-induced shock wave, given observations of sufficient resolution. 

For atmospheres and/or phases where no shock is present, the quasi-static approximation used to estimate stellar parameters, e.g. at maximum radius, has been shown wanting if the surface deceleration is an appreciable fraction of the surface gravity. It may lead to a considerable underestimate of the star's mean effective temperature and surface gravity.   

\section*{Acknowledgments}

The author is indebted to the following.
For motivation to keep working on ``the born-again rocket star'' -- the friends of V652 Herculis: Hideyuki, Pilar, Tony, Dave, Don, and Hiromoto. 
For motivation (a long time ago) to compute spectra for shocked atmospheres --  Myron Smith.
For making their computer programs available to wider communities -- Ulrich Heber and Detlef Sch\"onberner ({\sc sterne}), Philip Dufton ({\sc spectrum}), and Alan Bridger ({\sc nlpuls}).  
For a brilliant suggestion on how to plot shocks -- Laura Scott.
For financial support -- the UK Science and Technology Facilities Council via UKRI Grant No. ST/V000438/1 
 and the Northern Ireland Department for Communities which funds the Armagh Observatory and Planetarium.    
 For the purpose of open access, the author has applied a Creative Commons Attribution (CC BY) license to any Author Accepted Manuscript version arising.

\section*{Data Availability}

The models used as inputs for the current paper have been described previously and are available on reasonable request from the author. 
The computer programs used in the calculations are available upon reasonable request and at the user's risk. 
\bibliographystyle{mnras}
\bibliography{ehe}
\label{lastpage}
\end{document}

%% file: mydefs.tex
\newcommand{\Msolar}{\mbox{\,$\rm M_{\odot}$}}        
\newcommand{\Lsolar}{\mbox{\,$\rm L_{\odot}$}}        

  \newcommand{\Teff}{\mbox{\,\em T$_{\rm eff}$}}         
  \newcommand{\sg}{\mbox{\,log $g$}}                     

\newcommand{\nH}{\mbox{\,$n_{\rm H}$}}                 
  \newcommand{\kmsec}{\,\mbox{$\mbox{km}\,\mbox{s}^{-1}$}}    
  \def\simge{\mathrel{\raise1.16pt\hbox{$>$}\kern-7.0pt
    \lower3.06pt\hbox{{$\scriptstyle \sim$}}}}           
  \def\simle{\mathrel{\raise1.16pt\hbox{$<$}\kern-7.0pt
    \lower3.06pt\hbox{{$\scriptstyle \sim$}}}}           

%% file: v652_sim.bbl
\begin{thebibliography}{}
\makeatletter
\relax
\def\mn@urlcharsother{\let\do\@makeother \do\$\do\&\do\#\do\^\do\_\do\%\do\~}
\def\mn@doi{\begingroup\mn@urlcharsother \@ifnextchar [ {\mn@doi@}
  {\mn@doi@[]}}
\def\mn@doi@[#1]#2{\def\@tempa{#1}\ifx\@tempa\@empty \href
  {http://dx.doi.org/#2} {doi:#2}\else \href {http://dx.doi.org/#2} {#1}\fi
  \endgroup}
\def\mn@eprint#1#2{\mn@eprint@#1:#2::\@nil}
\def\mn@eprint@arXiv#1{\href {http://arxiv.org/abs/#1} {{\tt arXiv:#1}}}
\def\mn@eprint@dblp#1{\href {http://dblp.uni-trier.de/rec/bibtex/#1.xml}
  {dblp:#1}}
\def\mn@eprint@#1:#2:#3:#4\@nil{\def\@tempa {#1}\def\@tempb {#2}\def\@tempc
  {#3}\ifx \@tempc \@empty \let \@tempc \@tempb \let \@tempb \@tempa \fi \ifx
  \@tempb \@empty \def\@tempb {arXiv}\fi \@ifundefined
  {mn@eprint@\@tempb}{\@tempb:\@tempc}{\expandafter \expandafter \csname
  mn@eprint@\@tempb\endcsname \expandafter{\@tempc}}}

\bibitem[\protect\citeauthoryear{{Albrow} \& {Cottrell}}{{Albrow} \&
  {Cottrell}}{1994}]{albrow94}
{Albrow} M.~D.,  {Cottrell} P.~L.,  1994, \mn@doi [\mnras]
  {10.1093/mnras/267.3.548}, \href
  {https://ui.adsabs.harvard.edu/abs/1994MNRAS.267..548A} {267, 548}

\bibitem[\protect\citeauthoryear{{Alvarez}, {Jorissen}, {Plez}, {Gillet}  \&
  {Fokin}}{{Alvarez} et~al.}{2000}]{alvarez00}
{Alvarez} R.,  {Jorissen} A.,  {Plez} B.,  {Gillet} D.,   {Fokin} A.,  2000,
  \aap, \href {https://ui.adsabs.harvard.edu/abs/2000A&A...362..655A} {362,
  655}

\bibitem[\protect\citeauthoryear{{Behara} \& {Jeffery}}{{Behara} \&
  {Jeffery}}{2006}]{behara06}
{Behara} N.~T.,  {Jeffery} C.~S.,  2006, \aap, \href
  {http://ukads.nottingham.ac.uk/abs/2006A%26A...451..643B} {451, 643}

\bibitem[\protect\citeauthoryear{{B{\"o}hm} \& {Deinzer}}{{B{\"o}hm} \&
  {Deinzer}}{1965}]{bohm65}
{B{\"o}hm} K.~H.,  {Deinzer} W.,  1965, Zeitschrift fur Astrophysik, \href
  {http://ukads.nottingham.ac.uk/cgi-bin/nph-bib_query?bibcode=1965ZA.....61....1B&db_key=AST}
  {61, 1}

\bibitem[\protect\citeauthoryear{{Bridger}}{{Bridger}}{1983}]{bridger83}
{Bridger} A.,  1983, PhD thesis, University of St Andrews

\bibitem[\protect\citeauthoryear{{Byrne} \& {Jeffery}}{{Byrne} \&
  {Jeffery}}{2020}]{byrne20}
{Byrne} C.~M.,  {Jeffery} C.~S.,  2020, \mn@doi [\mnras]
  {10.1093/mnras/stz3486}, \href
  {https://ui.adsabs.harvard.edu/abs/2020MNRAS.492..232B} {492, 232}

\bibitem[\protect\citeauthoryear{{Chadid}, {Vernin}  \& {Gillet}}{{Chadid}
  et~al.}{2008}]{chadid08}
{Chadid} M.,  {Vernin} J.,   {Gillet} D.,  2008, \mn@doi [\aap]
  {10.1051/0004-6361:200810270}, \href
  {https://ui.adsabs.harvard.edu/abs/2008A&A...491..537C} {491, 537}

\bibitem[\protect\citeauthoryear{Christy}{Christy}{1967}]{christy67}
Christy R.~F.,  1967, Methods in Computational Physics, 7, 191

\bibitem[\protect\citeauthoryear{{Dufton}}{{Dufton}}{1972}]{dufton72}
{Dufton} P.~L.,  1972, \aap, \href
  {http://ukads.nottingham.ac.uk/abs/1972A%26A....16..301D} {16, 301}

\bibitem[\protect\citeauthoryear{Fadeyev \& Lynas-Gray}{Fadeyev \&
  Lynas-Gray}{1996}]{fadeyev96}
Fadeyev Y.~A.,  Lynas-Gray A.~E.,  1996, \mnras, \href
  {http://adsabs.harvard.edu/abs/1996MNRAS.280..427F} {280, 427}

\bibitem[\protect\citeauthoryear{{Fokin}}{{Fokin}}{1990}]{fokin90}
{Fokin} A.,  1990, \mn@doi [\apss] {10.1007/BF00653554}, \href
  {https://ui.adsabs.harvard.edu/abs/1990Ap&SS.164...95F} {164, 95}

\bibitem[\protect\citeauthoryear{{Fokin}, {Mathias}, {Chapellier}, {Gillet}  \&
  {Nardetto}}{{Fokin} et~al.}{2004}]{fokin04}
{Fokin} A.,  {Mathias} P.,  {Chapellier} E.,  {Gillet} D.,   {Nardetto} N.,
  2004, \mn@doi [\aap] {10.1051/0004-6361:20040418}, \href
  {https://ui.adsabs.harvard.edu/abs/2004A&A...426..687F} {426, 687}

\bibitem[\protect\citeauthoryear{{Heber}, {Hunger}, {Jonas}  \&
  {Kudritzki}}{{Heber} et~al.}{1984}]{heber84}
{Heber} U.,  {Hunger} K.,  {Jonas} G.,   {Kudritzki} R.~P.,  1984, \aap, \href
  {http://ukads.nottingham.ac.uk/cgi-bin/nph-bib_query?bibcode=1984A%26A...130..119H&db_key=AST}
  {130, 119}

\bibitem[\protect\citeauthoryear{Hill, Kilkenny, Sch\"{o}nberner  \&
  Walker}{Hill et~al.}{1981}]{hill81}
Hill P.~W.,  Kilkenny D.,  Sch\"{o}nberner D.,   Walker H.~J.,  1981, \mnras,
  \href {http://adsabs.harvard.edu/abs/1981MNRAS.197...81H} {197, 81}

\bibitem[\protect\citeauthoryear{{Hunger} \& {van Blerkom}}{{Hunger} \& {van
  Blerkom}}{1967}]{hunger67}
{Hunger} K.,  {van Blerkom} D.,  1967, \zap, \href
  {https://ui.adsabs.harvard.edu/abs/1967ZA.....66..185H} {66, 185}

\bibitem[\protect\citeauthoryear{{Jeffery} \& {Heber}}{{Jeffery} \&
  {Heber}}{1992}]{jeffery92}
{Jeffery} C.~S.,  {Heber} U.,  1992, \aap, \href
  {http://ukads.nottingham.ac.uk/abs/1992A%26A...260..133J} {260, 133}

\bibitem[\protect\citeauthoryear{{Jeffery} \& {Hill}}{{Jeffery} \&
  {Hill}}{1986}]{jeffery86}
{Jeffery} C.~S.,  {Hill} P.~W.,  1986, \mnras, \href
  {http://cdsads.u-strasbg.fr/abs/1986MNRAS.221..975J} {221, 975}

\bibitem[\protect\citeauthoryear{{Jeffery}, {Woolf}  \& {Pollacco}}{{Jeffery}
  et~al.}{2001}]{jeffery01b}
{Jeffery} C.~S.,  {Woolf} V.~M.,   {Pollacco} D.~L.,  2001, \aap, \href
  {http://ukads.nottingham.ac.uk/abs/2001A%26A...376..497J} {376, 497}

\bibitem[\protect\citeauthoryear{{Jeffery}, {Shibahashi}, {Kurtz}, {Elkin},
  {Monta{\~n}{\'e}s-Rodr{\'\i}guez}  \& {Saio}}{{Jeffery}
  et~al.}{2013}]{jeffery13.fuji2}
{Jeffery} C.~S.,  {Shibahashi} H.,  {Kurtz} D.~W.,  {Elkin} V.,
  {Monta{\~n}{\'e}s-Rodr{\'\i}guez} P.,   {Saio} H.,  2013, in {Shibahashi} H.,
   {Lynas-Gray} A.~E.,  eds,  Astronomical Society of the Pacific Conference
  Series Vol. 479, Progress in Physics of the Sun and Stars: A New Era in
  Helio- and Asteroseismology. p.~369

\bibitem[\protect\citeauthoryear{{Jeffery}, {Kurtz}, {Shibahashi}, {Starling},
  {Elkin}, {Monta{\~n}{\'e}s-Rodr{\'{\i}}guez}  \& {McCormac}}{{Jeffery}
  et~al.}{2015}]{jeffery15b}
{Jeffery} C.~S.,  {Kurtz} D.,  {Shibahashi} H.,  {Starling} R.~L.~C.,  {Elkin}
  V.,  {Monta{\~n}{\'e}s-Rodr{\'{\i}}guez} P.,   {McCormac} J.,  2015, \mn@doi
  [\mnras] {10.1093/mnras/stu2654}, \href
  {http://adsabs.harvard.edu/abs/2015MNRAS.447.2836J} {447, 2836}

\bibitem[\protect\citeauthoryear{{Jeffery}, {Monta{\~n}{\'e}s-Rodr{\'\i}guez}
  \& {Saio}}{{Jeffery} et~al.}{2022}]{jeffery22a}
{Jeffery} C.~S.,  {Monta{\~n}{\'e}s-Rodr{\'\i}guez} P.,   {Saio} H.,  2022,
  \mn@doi [\mnras] {10.1093/mnras/stab2876}, \href
  {https://ui.adsabs.harvard.edu/abs/2022MNRAS.509.1940J} {509, 1940}

\bibitem[\protect\citeauthoryear{{Joy}}{{Joy}}{1947}]{joy47}
{Joy} A.~H.,  1947, \mn@doi [\apj] {10.1086/144959}, \href
  {https://ui.adsabs.harvard.edu/abs/1947ApJ...106..288J} {106, 288}

\bibitem[\protect\citeauthoryear{{Kilkenny}, {Koen}, {Jeffery}, {Hill}  \&
  {O'Donoghue}}{{Kilkenny} et~al.}{1999}]{kilkenny99}
{Kilkenny} D.,  {Koen} C.,  {Jeffery} C.~S.,  {Hill} N.~C.,   {O'Donoghue} D.,
  1999, \mnras, \href {http://ukads.nottingham.ac.uk/abs/1999MNRAS.310.1119K}
  {310, 1119}

\bibitem[\protect\citeauthoryear{{Kurucz}}{{Kurucz}}{1979}]{kurucz79}
{Kurucz} R.~L.,  1979, \mn@doi [\apjs] {10.1086/190589}, \href
  {https://ui.adsabs.harvard.edu/abs/1979ApJS...40....1K} {40, 1}

\bibitem[\protect\citeauthoryear{{Kurucz}}{{Kurucz}}{2005}]{kurucz05}
{Kurucz} R.~L.,  2005, Memorie della Societa Astronomica Italiana Supplementi,
  \href {https://ui.adsabs.harvard.edu/abs/2005MSAIS...8...14K} {8, 14}

\bibitem[\protect\citeauthoryear{{Lawson}}{{Lawson}}{1986}]{lawson86}
{Lawson} W.~A.,  1986, in {Hunger} K.,  {Schoenberner} D.,   {Kameswara Rao}
  N.,  eds, IAU Colloq. 87: Hydrogen Deficient Stars and Related Objects.
  Reidel, p.~211, \mn@doi{10.1007/978-94-009-4744-3\_20}

\bibitem[\protect\citeauthoryear{{Lynas-Gray}, {Sch\"{o}nberner}, {Hill}  \&
  {Heber}}{{Lynas-Gray} et~al.}{1984}]{lynasgray84}
{Lynas-Gray} A.~E.,  {Sch\"{o}nberner} D.,  {Hill} P.~W.,   {Heber} U.,  1984,
  \mnras, \href {http://ukads.nottingham.ac.uk/abs/1984MNRAS.209..387L} {209,
  387}

\bibitem[\protect\citeauthoryear{{Martin}}{{Martin}}{2019}]{martin19.phd}
{Martin} P.,  2019, PhD thesis, Trinity College Dublin

\bibitem[\protect\citeauthoryear{{Monta{\~n}{\'e}s Rodriguez} \&
  {Jeffery}}{{Monta{\~n}{\'e}s Rodriguez} \& {Jeffery}}{2001}]{montanes01}
{Monta{\~n}{\'e}s Rodriguez} P.,  {Jeffery} C.~S.,  2001, \mn@doi [\aap]
  {10.1051/0004-6361:20010840}, \href
  {http://adsabs.harvard.edu/abs/2001A%26A...375..411M} {375, 411}

\bibitem[\protect\citeauthoryear{{Monta{\~n}{\'e}s Rodr{\'\i}guez} \&
  {Jeffery}}{{Monta{\~n}{\'e}s Rodr{\'\i}guez} \& {Jeffery}}{2002}]{montanes02}
{Monta{\~n}{\'e}s Rodr{\'\i}guez} P.,  {Jeffery} C.~S.,  2002, \mn@doi [\aap]
  {10.1051/0004-6361:20020019}, \href
  {https://ui.adsabs.harvard.edu/abs/2002A&A...384..433M} {384, 433}

\bibitem[\protect\citeauthoryear{{Odgers}}{{Odgers}}{1956}]{odgers56}
{Odgers} G.~J.,  1956, Publications of the Dominion Astrophysical Observatory
  Victoria, \href {https://ui.adsabs.harvard.edu/abs/1956PDAO...10..215O} {10,
  215}

\bibitem[\protect\citeauthoryear{{Parsons}}{{Parsons}}{1972}]{parsons72}
{Parsons} S.~B.,  1972, \mn@doi [\apj] {10.1086/151468}, \href
  {http://adsabs.harvard.edu/abs/1972ApJ...174...57P} {174, 57}

\bibitem[\protect\citeauthoryear{{Przybilla}, {Butler}, {Heber}  \&
  {Jeffery}}{{Przybilla} et~al.}{2005}]{przybilla05}
{Przybilla} N.,  {Butler} K.,  {Heber} U.,   {Jeffery} C.~S.,  2005, \aap,
  \href {http://ukads.nottingham.ac.uk/abs/2005A%26A...443L..25P} {443, L25}

\bibitem[\protect\citeauthoryear{{Sanford}}{{Sanford}}{1949}]{sanford49}
{Sanford} R.~F.,  1949, \mn@doi [\pasp] {10.1086/126151}, \href
  {https://ui.adsabs.harvard.edu/abs/1949PASP...61..135S} {61, 135}

\bibitem[\protect\citeauthoryear{{Sasselov} \& {Raga}}{{Sasselov} \&
  {Raga}}{1992}]{sasselov92}
{Sasselov} D.~D.,  {Raga} A.,  1992, in {Giampapa} M.~S.,  {Bookbinder} J.~A.,
  eds,  Astronomical Society of the Pacific Conference Series Vol. 26, Cool
  Stars, Stellar Systems, and the Sun. p.~549

\bibitem[\protect\citeauthoryear{{Schwarzschild}}{{Schwarzschild}}{1952}]{schwarzschild52}
{Schwarzschild} M.,  1952, Trans. IAU, VIII, 811

\bibitem[\protect\citeauthoryear{{Seaton}, {Yan}, {Mihalas}  \&
  {Pradhan}}{{Seaton} et~al.}{1994}]{OP94}
{Seaton} M.~J.,  {Yan} Y.,  {Mihalas} D.,   {Pradhan} A.~K.,  1994, \mnras,
  \href {http://adsabs.harvard.edu/abs/1994MNRAS.266..805S} {266, 805}

\bibitem[\protect\citeauthoryear{{Wallerstein}}{{Wallerstein}}{1959}]{wallerstein59}
{Wallerstein} G.,  1959, \mn@doi [\apj] {10.1086/146745}, \href
  {https://ui.adsabs.harvard.edu/abs/1959ApJ...130..560W} {130, 560}

\bibitem[\protect\citeauthoryear{{Whitney}}{{Whitney}}{1956}]{whitney56}
{Whitney} C.,  1956, \mn@doi [\aj] {10.1086/107324}, \href
  {https://ui.adsabs.harvard.edu/abs/1956AJ.....61..192W} {61, 192}

\bibitem[\protect\citeauthoryear{{Wolf}}{{Wolf}}{1973}]{wolf73}
{Wolf} R.~E.~A.,  1973, \aap, \href
  {https://ui.adsabs.harvard.edu/abs/1973A&A....26..127W} {26, 127}

\bibitem[\protect\citeauthoryear{{Woolf} \& {Jeffery}}{{Woolf} \&
  {Jeffery}}{2002}]{woolf02}
{Woolf} V.~M.,  {Jeffery} C.~S.,  2002, \mn@doi [\aap]
  {10.1051/0004-6361:20021113}, \href
  {http://cdsads.u-strasbg.fr/abs/2002A%26A...395..535W} {395, 535}

\bibitem[\protect\citeauthoryear{\protect{The Opacity Project
  Team}}{\protect{The Opacity Project Team}}{1995}]{OP95}
\protect{The Opacity Project Team} 1995, The Opacity Project Vol. 1.
Institute of Physics Publications, Bristol, UK

\makeatother
\end{thebibliography}
